\documentclass[referee]{aa}
\usepackage{graphics}
\usepackage{longtable}
\usepackage{psfig}
\usepackage{astron}

\newcommand{\bmv}{(B-V)}
\newcommand{\vmi}{(V-I)}
\newcommand{\umb}{(U-B)}
\begin{document}

   \thesaurus{05(08.08.1;           
		 08.08.2;	    
		 08.12.3;	    
		 08.16.3;	    
		 10.07.3 NGC 2808)} 
%
    \title{The anomalous Galactic globular cluster NGC 2808.\thanks{ 
Based on observations made at the European Southern
Observatory, La Silla, Chile, and on observations with the NASA/ESA
{\it Hubble Space Telescope}.
		}
	}
    \subtitle{Mosaic wide-field multi-band photometry. }

    \author{L. R. Bedin\inst{1}, G. Piotto\inst{1}, M. Zoccali\inst{1},
	P. B. Stetson\inst{2}, I. Saviane\inst{1}, S. Cassisi\inst{3}, and G. Bono\inst{4}}

    \offprints{G. Piotto}

    \mail{piotto@pd.astro.it}

    \institute{ Dipartimento di Astronomia, Universit\`a di Padova, Vicolo dell'Osservatorio 5, I-35122 Padova, Italy \and 
	   Dominion Astrophysical Observatory, Herzberg Institute of Astrophysics, National Research Council, 5071 West Saanich Road, Victoria, BC, Canada \and 
           Osservatorio Astronomico di Collurania, Via M. Maggini, I-64100 Teramo, Italy \and
	   Osservatorio Astronomico di Roma, Via Frascati 33, Monteporzio Catone, I-00040 Roma, Italy
	  }

    \date{Received 12 April 2000 / Accepted 20 June 2000}

   \maketitle

   \markboth{Bedin et al.}{The anomalous NGC 2808}

   \begin{abstract} We present Johnson $UBV$, and Kron-Cousins $I$
   photometry for about 60,000 stars in a region of $44\times33$
   arcmin$^2$ centered on the Galactic globular cluster (GGC) NGC 
   2808. The central $r<100$ arcsec region has been mapped also in the
   F218W, F439W, and F555W bands with the $HST$ WFPC2 camera.
   Overall, we cover a field which extends from the center to about
   1.7 tidal radii. Photometric calibration has been secured by more
   than 1000 standards, and the $HST$ $B$ and $V$ photometry has been
   linked to the groundbased system.

We confirm the anomalously elongated horizontal bra\-nch (HB) with gaps
along it, and we show that the gaps are statistically significant.
Most importantly, we find that:
\begin{itemize}
\item
the extended blue tail of HB (EBT) is present beyond $400$ arcsec,
corresponding to more than nine times the half mass radius $r_{\rm h}$, and
extends to $V=21.2$ also in these external regions;
\item
also the gaps on the EBT are present at least out to 400 arcsec from
the cluster center, and possibly beyond that.  The location of the
gaps in the HB seems to be the same all over the cluster.
\item 
there are no significant radial gradients in the distribution of the
stars in the different HB stumps and some marginal evidence of an
increase in the number of red giant branch (RGB) stars in the cluster
core.
\end{itemize}
The observational facts presented in this paper seem to exclude the
possibility that the EBTs originate from tidal stripping of the
envelope of the RGB stars due to close encounters in high density
environments or to mass transfer in close binaries.  Also, it is not
clear whether the presence of EBTs and gaps on them are the
manifestation of the same physical phenomenon.

We have also shown that the jump recently identified in the HB using
Str\"omgren-$u$ photometry is clearly visible also in the $U$, $U-B$
plane, at $T_{\rm e}\sim11,600$ K.

Finally, we present the luminosity function (LF) of the RGB. The LF
clearly shows the presence of the usual RGB bump, but also a second
feature, 1.4 magnitudes brighter in $V$, that we have named RGB {\it
heap}. The RGB heap, visible also in other GGCs, is a new feature
that, because of its position on the RGB, we have tentatively
associated with the recently discovered K giant variables.
\end{abstract}


\section{Introduction\label{intro}}
After more than three decades of observational and theoretical
efforts, the causes of the star distribution along the horizontal
branch (HB) in Galactic globular clusters (GGC) remain obscure. In
general, the HBs tend to become bluer and bluer with decreasing
metallicity. However, there is a large variety of HB morphologies, and
many HBs do not have the color expected for their metal content.
Among these, a special case is represented by those clusters that have
extended-HB blue tails (EBT), indicating that some of the HB stars
lost almost all of their envelope during the red giant branch (RGB)
phase or at the helium core flash (Ferraro et al. 1998; Piotto et
al. 1999, and references therein). In many GGCs there is another
peculiarity: a bimodal or multimodal distribution along the HB
(Catelan et al. 1998), which sometimes results in a gap, i.e., a
region clearly underpopulated in stars. All the EBTs present at least
one gap along them (Ferraro et al. 1998; Piotto et al. 1999).
No clear explanation for the origin of the gaps is available at
present.

Surely, NGC 2808 represents the most extreme known example of these
anomalous GCs. Since Harris \cite*{harr74}, it is known that its HB is
rather unusual. In fact, it is rich in both red and blue HB stars, but
almost completely lacking in intermediate color objects. Despite the
richness of the total HB population, Clement \& Hazen 
\cite*{clem89} found only two RR-Lyrae variables.
The bimodality of the HB has been confirmed in many subsequent color
magnitude diagram (CMD) investigations of NGC 2808 (Walker 1999,
and references therein, hereafter W99), and in some other clusters
(Catelan et al. 1998)
However, only the deep CMD in the $B$, $V$, and UV (F218W) bands ---
based on $HST$ WFPC2 data --- allowed the discovery of the other
anomalies which make NGC 2808 probably a unique case (Sosin et
al. 1997,
hereafter S97). S97 showed that the blue HB extends down to very faint
magnitudes, well below the main sequence (MS) turn-off. Most
importantly, the EBT has also two narrow gaps.  S97 could not find any
plausible explanation either for the HB
extension or for the multimodal distribution of the HB stars. As
already suggested by Rood et al. (1993),
S97 excluded that these could be due
to an age or metallicity dispersion. Also merging events can be
excluded (Catelan et al. 1998; S97). No other plausible explanations for
the observed HB have been found, yet.

The discovery of the peculiar HB in NGC 2808 is part of a long term
$HST$ program aimed at identifying anomalous stellar populations in
the cores of GGCs (Piotto et al. 1999). 
As groundbased followup, we are also carrying out a project for a wide
field mapping of the envelopes of the clusters observed with $HST$. It
has been often suggested that dynamical interactions in dense clusters
could modify their stellar populations (Djorgovski \& Piotto 1993;
Fusi Pecci et al. 1993). A combination of the high
resolution $HST$ images in the densest inner region of a GGC with the
wide field frames of the outer parts allows a direct test of this
hypothesis. In this respect, the extreme properties of NGC 2808 offer
a unique opportunity. Surely it is of interest to see whether the HB
maintains its morphology also in the outskirts of the cluster.

In this paper we present wide-field stellar photometry in $UBVI$ of
the entire cluster, extending beyond its tidal radius. While we were
working to this project, a wide field $BV$ CMD diagram of NGC 2808 has
also been published by W99.  Our study complements the W99 work,
extending the photometry to stars in outer regions, and to the $U$ and
$I$ bands. A list of the main cluster parameters can be found in
Table~\ref{ph_tab}.
   \begin{table}
      \caption[]{The main parameters of NGC 2808: Right Ascension, declination, 
Galactic longitude, Galactic latitude, distance from sun, distance to Galactic 
Center, distance to Galactic plane, metallicity, reddening, apparent distance 
modulus, core, half-mass, tidal, radii, concentration parameter, 
logarithm of core relaxation time and at half-mass radius 
in $\log_{10}$(years). }
\label{ph_tab}
      \[
         \begin{array}{p{0.5\linewidth}l}
            \hline
            \noalign{\smallskip}
            Parameter     &  Value  \\
            \noalign{\smallskip}
            \hline
            \noalign{\smallskip}
            	R.A.$^{\rm a}$ (J2000)    	& 09^{\rm h}12^{\rm m}02^{\rm s}.6\\
            	$\delta^{\rm a}$ (J2000)    	& -64^{o}51'47''      	\\
            	$l_{\rm II}^{\rm a}$        	& 282^{o}.192        	\\
            	$b_{\rm II}^{\rm a}$        	& -11^{o}.253	     	\\
	    	$R_{\odot}^{\rm b}$         	&  9.3 ~ {\rm Kpc} 	\\
            	$R_{\rm GC}^{\rm b}$		& 11.0 ~ {\rm Kpc}	\\
            	$R_{\rm GP}^{\rm b}$	        & -1.8 ~ {\rm Kpc}	\\
		${\rm [Fe/H]}^{\rm c}$ 		& -1.24			\\
		$E_{B-V}^{\rm d}$		&  0.19			\\
		$(m-M)_{V}^{\rm d}$		& 15.79			\\
		$ r_{\rm c}^{\rm b}$		&  0'.26  		\\
		$ r_{\rm h}^{\rm b}$		&  0'.76  		\\
		$ r_{\rm t}^{\rm b}$		&  15'.55		\\
		$ c^{\rm b} $		        &  1.77			\\
		$Log(t_{\rm c})^{\rm b}$ [year]	&  8.28  		\\
		$Log(t_{\rm h})^{\rm b}$ [year]	&  9.11			\\
            \noalign{\smallskip}
            \hline
         \end{array}
      \]
\begin{list}{}{}
\item[$^{\mathrm{a}}$] from Djorgovski \& Meylan \cite*{djor93b}
\item[$^{\mathrm{b}}$] from Harris
(http://physun.physics.\-mcmaster.ca/GC//\-mwgc.dat; June 22, 1999
revison)
\item[$^{\mathrm{c}}$] from Carretta \& Gratton \cite*{carr97} scale as extended by
Cohen et al. \cite*{cohe99}
\item[$^{\mathrm{d}}$] this work (cf. Sect.~\ref{obsR})
\end{list}
   \end{table}

\section{Observations and reduction\label{obs}}
\subsection{The data set \label{data}}
NGC 2808 has been observed in the $UBVI$ bands in January 1998 with
DFOSC mounted on the 1.54 ESO-$Danish$ telescope at La Silla. DFOSC
was equipped with a 2048 $\times$ 2048 thinned LORAL CCD, with a pixel
scale of 0.39$''$/pix\-el. An area of $\sim44\times33$ arcmin$^2$
(Fig.\ref{danf}) has been mapped in $BV$ with a set of 9 partially
overlapping fields. We took two $B$ and two $V$ exposures per field
(50, 1800s in $B$, and 25, 900s in $V$), plus an additional 25s
exposure in $I$ for calibration purposes. An additional
12.9$\times$12.9 $arcmin^2$ field centered on the center of NGC 2808
has been covered with $300+5\times1200$s in $U$, $30+900$s in $B$,
420s in $V$, and 25s in $I$.  Due to the combined effects of seeing
and charge bleeding, stellar images had a FWHM around 1.5 arcsec in
all the images, but the nights were all photometric.
\begin{figure}
{\par\centering \resizebox*{1\columnwidth}{!}{\includegraphics{H2150.f01}} \par}
\caption{The digital mosaic of the 10 $\sim12.9\times12.9$ ${\rm arcmin}^2$  
CCD images collected at the ESO-$Danish$ telescope. North is up and East is left. } 
\label{danf}
\end{figure}
NGC 2808 has also been observed in the $VI$ bands in March 1995 with
EMMI mounted on the ESO-$NTT$ telescope. An area of $\sim 30\times36$
arcmin$^2$ (Fig.\ref{nttf}) has been covered with a mosaic of twelve
$7.5\times7.5$ arcmin$^2$ partially overlapping fields. For each field
we obtained one 25s exposure both in $V$ and $I$.
An additional field centered at $\sim7$ arcmin from the cluster
center has been covered with exposures of $3\times600s$ in $V$
$+~5\times600s$ in $I$.
The seeing was stable and around 0.9 arcsec (FWHM), but the night was not photometric.
\begin{figure}
{\par\centering \resizebox*{1\columnwidth}{!}{\includegraphics{H2150.f02}} \par}
\caption{The digital mosaic of the 13 $\sim7.5\times7.5$ arcmin$^2$
CCD images collected at the ESO-$NTT$ telescope. North is up and East is left.\label{nttf}}
\end{figure}
Finally, we also re-reduced the F439W and F555W WFPC2 $HST$ images
used by S97 in their investigation. In this case, the total exposure
time was 510s and 57s in F439W and F555W bands, respectively.  As the
PC camera was approximately centered on the center of NGC 2808, the
WFPC2 data allow us to investigate the inner $\sim100$ arcsec of the
cluster, while the groundbased mosaics extend to $\sim1.7 r_{\rm t}$, where
$r_{\rm t}$ is the tidal radius ($r_{\rm t}=15'.6$, Table~\ref{ph_tab}).
\subsection{Data reduction}
All the groundbased images have been pre-processed in the standard way
with IRAF, using the sets of bias and sky flat-field images collected
during the observing nights. The PSFs have been obtained with the
standard routines in DAOPHOT II (Stetson 1987), 
while we used ALLFRAME (Stetson 1994)
on all the images for the actual PSF-fitting photometry.
The $HST$ data have been reduced again (after S97). In S97, for the
stellar photometry, they used ALLSTAR (Stetson 1993)
and a set of PSFs
derived directly from the single NGC 2808 images.  Here, we used
ALLFRAME on a new list obtained from a median image of all frames
(Stetson 1994).  The new reduction allowed us to extend the photometry
from the $HST$ data by $\sim1$ magnitude fainter than in S97. This is an
important improvement, in particular for the determination of the
faintest limit of the EBT (cf. Sect.~\ref{hb}).
\subsection{Calibration to a standard photometric system\label{cal}}
Particular care has been devoted to link the instrumental magnitudes (from both
the groundbased and $HST$ data) to a photometric standard system.
\subsubsection{Groundbased data}
Only the nights at the $Danish$ telescope were photometric. In order
to obtain the transformation equations relating the instrumental
($ubvi$) magnitudes to the standard $UBV$(Johnson), $I$(Kron-Cousins)
system, six Landolt (Landolt 1992)
fields of standards have been
observed. Specifically: the selected areas 95, 98, and the Rubin 149,
PG0918+029, PG1047+003, and PG1323-086 fields. In each of these, there
are other secondary standard stars by Stetson \cite*{stet00} which
extend the previous Landolt sequence to 90, 618, 147, 80, 45, and 63
standards, respectively, for each field. Moreover, 34 stars in the
field of NGC$\,$2808 from Harris (Harris 1978) 
and 19 from
Walker\footnote{ Note that we used the correct stars listed in erratum 
of Walker \protect\cite*{walk00}.  } (\-W\-99) were used as additional
photometric standards for this calibration. For all 1096 of these
stars aperture photometry was obtained on all the images after the
digital subtraction of neighboring objects from the frames. Most
importantly, these standard stars cover the color interval: $ -0.3 <
\bmv < 2.0 $.  We used DAOGROW (Stetson 1990) 
to obtain the aperture
growth curve for every frame and other related codes (by PBS) to
determine the aperture corrections, the transformation coefficients,
and the final calibrated photometry (e.g., Stetson 1993).  We
used transformation equations of the form 
$$ v = V + A_0 + A_1 * X + A_2 * \bmv,$$ 
$$ b = B + B_0 + B_1 * X + B_2 * \bmv,$$ 
$$ i = I + C_0 + C_1 * X + C_2 * \vmi,$$ 
$$ u = U + E_0 + E_1 * X + E_2 * \umb.$$ 
Second order color terms were tried and turned out to be negligible in
comparison to their uncertainties in all four filters. It is a
reasonable hypothesis that the color-dependent terms imposed by the
telescope, filter, and detector should be reasonably constant over the
course of a few nights, so after $A_2$, $B_2$, and $C_2$ had been
independently determined from each night's data, we obtained weighted
mean values of these parameters and imposed them as fixed quantities
for the definitive calibration; note that we had only one night of
observations in $U$ band.  The NGC$\,$2808 images enabled us to define
1690 local standards that met the following conditions: each was well
separated from its neighbors (according to a criterion described in
Stetson \cite*{stet93} Sect. 4.1), each was observed at least eight times,
each had a standard error of the mean magnitude (averaged over all
filters) less than 0.01~mag, each had a mean instrumental magnitude
less than 14.5 (again averaged over all filters), each had a mean
value of the goodness-of-fit statistic, CHI, less than 2, and each had
a mean value of the image-shape statistics, SHARP, between $-0.5$ and
$+0.5$. Once calibrated, this local standard sequence enables us to
place each of the photometric and nonphotometric observations of NGC
2808 at the $Danish$ telescope on the same magnitude system with an
uncertainty of $0.002$ magnitudes for $V$, $0.003$ for $B$, $0.028$
for $I$ and $0.001$ for $U$ (typical uncertainty in the relative zero
point of one frame as referred to the average of all frames). In a
similar way, the $NTT$ data have been calibrated to the average
$Danish$ photometric system above defined with an uncertainty of 0.003
magnitudes in $V$ and 0.004 magnitude in $I$.
\subsubsection{$HST$ calibration}
Calibration of the $HST$ data of S97 --- even updated with the most recent
Charge Transfert Efficiency correction by
Wiggs et al. \cite*{wigg97} 
--- shows a strong disagreement (cf. Fig.\ref{fid_cmd}) with our and
previous W99 photometry, being almost 0.1 mag bluer in $(B-V)$ than
the ground-based results. Therefore, to obtain a calibration as
homogeneus as possible between the two samples, the $HST$ data have
been calibrated by comparing the magnitudes of a number of convenient
common stars to the $Danish$ ground-based data producing a linear
transformation between $HST$ data and $Danish$ data.  These comparison
stars had to satisfy the following criteria: each had at least a
radial distance of $59$ arcsec from the estimated position of the center
of the cluster, each had a $Danish$ magnitude brighter than 18.00 in
$B$, each had a ground-based error less than 0.20 in $B$, and each had
an $HST$ error less than 0.05 in $B_{HST}$. Among these a $k$-$\sigma$
clipping procedure were followed to choose the best mean magnitude
difference based on 164 stars in $B$ and 115 in $V$.

After this procedure, the $B$ and $V$ magnitudes for the $HST$ data
were linked to the groundbased $Danish$ system previously defined
with an uncertainty of 0.003 magnitudes in both the two bands. 
\section{The Color Magnitude Diagrams}\label{obscmd}
\begin{figure}
{\par\centering \resizebox*{1\columnwidth}{!}{\includegraphics{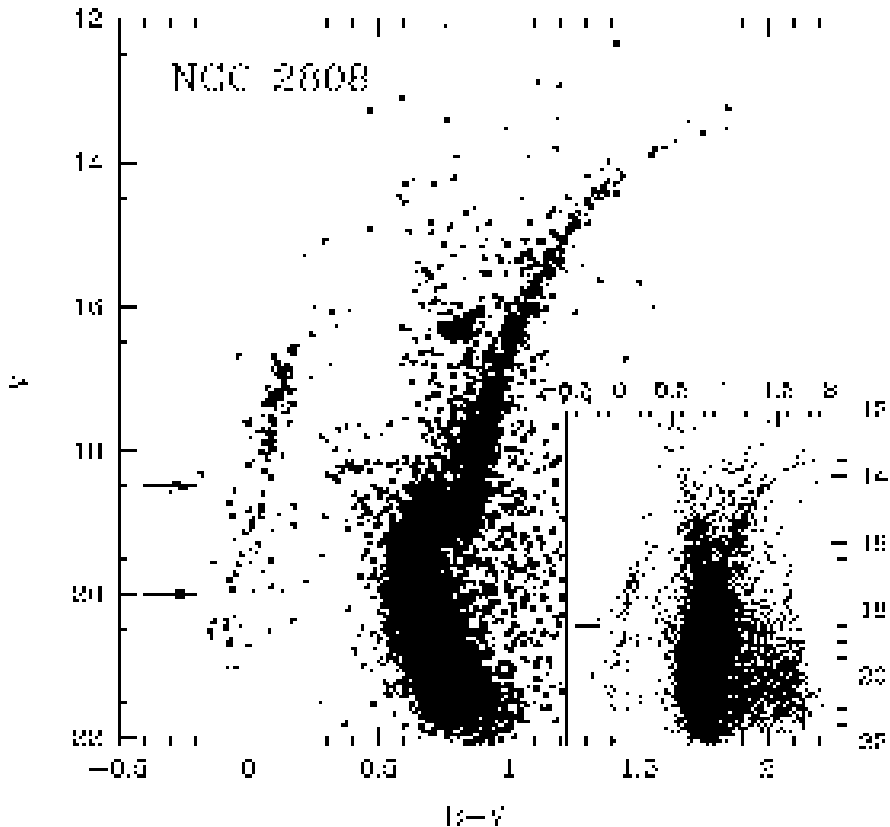}} \par}
\caption{The $V$ vs. ($B-V$) CMD from the $Danish$ data. The main figure shows the
14,100 stars with $100<r<400$ arcsec and photometric error $<$ 0.05; the CMD in the figure
inset corresponds to the 30,000 stars with $100<r<r_{\rm t}$ arcsec and the same selection. The
two arrows show the location of the two gaps in the HB EBT (see text). \label{cmd_bv_dan}}
\end{figure}

\begin{figure}
{\par\centering \resizebox*{1\columnwidth}{!}{\includegraphics{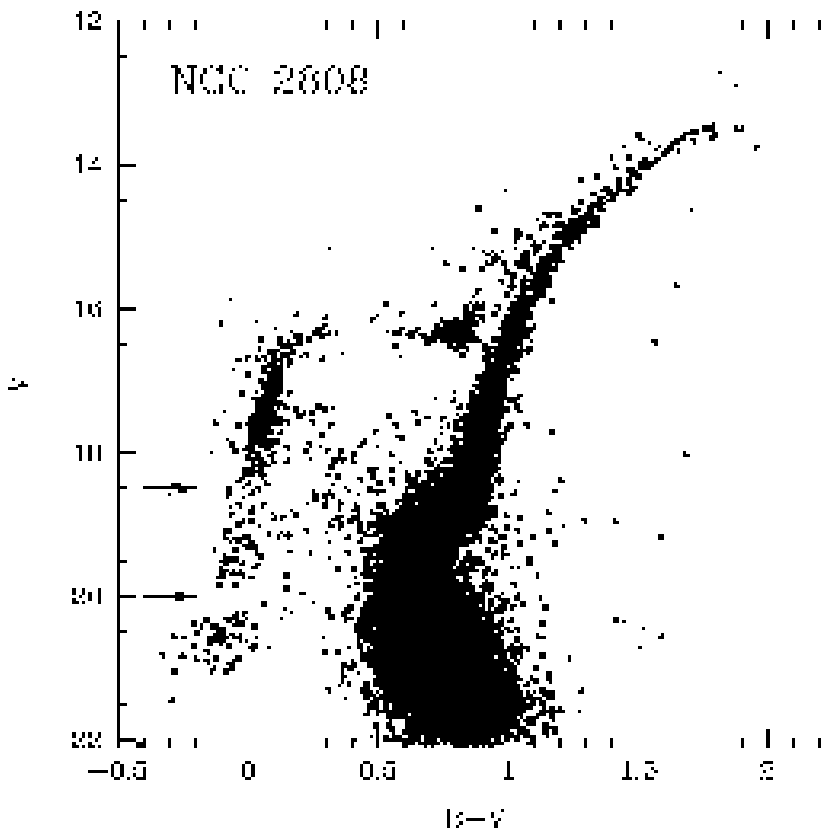}} \par}
\caption{The $V$ vs. ($B-V$) CMD from the $HST$ WFPC2 data of the inner core of NGC 2808
($r\leq100$ arcsec). Only the 35,000 stars with a photometric
error less than 0.15 are plotted.\label{cmd_hst}}
\end{figure}

\begin{figure}
{\par\centering \resizebox*{1\columnwidth}{!}{\includegraphics{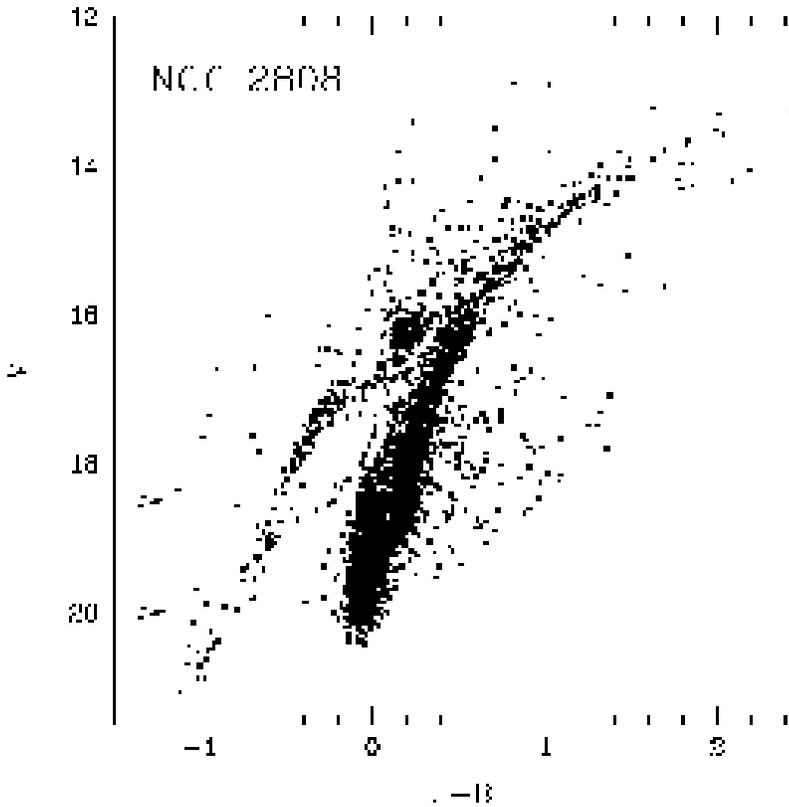}} \par}
\caption{The $V$ vs. ($U-B$) CMD for 6,000 stars from the $Danish$ data
with error $<$ 0.03.  We have covered only one 800$\times$800
arcsec$^2$ field in the $U$-band.\label{cmd_ub}}
\end{figure}

\begin{figure}
{\par\centering \resizebox*{1\columnwidth}{!}{\includegraphics{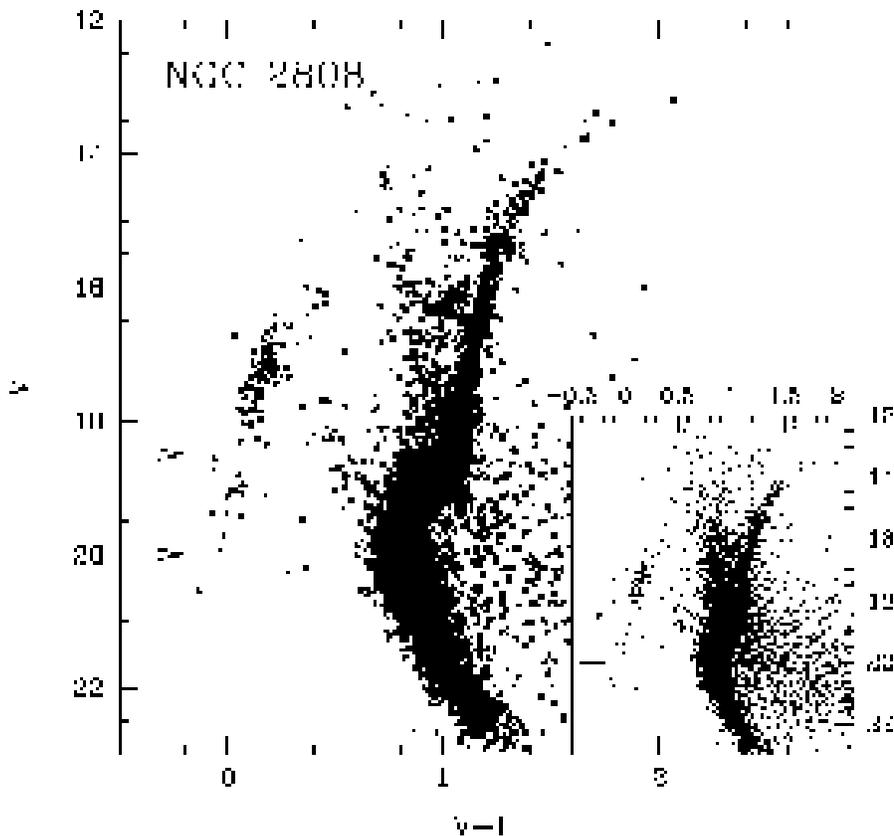}} \par}
\caption{As in Fig.~\ref{cmd_bv_dan}, but in the $V$ vs. ($V-I$)
plane. In the large figure 12,000 stars are showed; 17,700 in the inset. }
\label{cmd_vi}
\end{figure}

The CMDs derived from the photometry of about 60,000 stars discussed
in the previous section are presented in
Figs.~\ref{cmd_bv_dan}$-$\ref{cmd_vi}.  This is the most complete
photometry of the NGC 2808 stellar population published so far. It
complements the CMDs presented by S97 and W99 both by extending the
radial coverage to the tidal radius (and beyond) and by adding the $U$
and $I$ photometry.  The $V$ vs. ($B-V$) CMDs are presented in
Fig.~\ref{cmd_bv_dan} (groundbased data), and Fig.~\ref{cmd_hst}
($HST$ data).  The larger CMD in Fig.~~\ref{cmd_bv_dan} shows the
stars with $100<r<400$ arcsec, while the CMD in the inset shows all
the 30,000 stars with $100<r<r_{\rm t}$ arcsec and a photometric error $<$
0.05.  Note how the CMD of Fig.~\ref{cmd_hst} goes $\sim1$ magnitude
fainter than in S97, enabling us to cover the entire HB even in the
very inner regions ($r\leq 100$ arcsec). The groundbased and the $HST$
data have similar limiting magnitudes at $V\sim22$. Fig.~\ref{cmd_ub}
shows the CMD in the $V$ vs. ($U-B$) plane ($Danish$ data only), while
Fig.~\ref{cmd_vi} shows the $V$ vs. ($V-I$) diagram (from the
combination of the $Danish$ and $NTT$ data). The two diagrams of
Fig.~\ref{cmd_vi} refer to the same annuli defined in
Fig.~\ref{cmd_bv_dan}.  The CMDs of
Figs.~\ref{cmd_bv_dan}$-$\ref{cmd_vi} clearly show both the
extended HB tails and the gaps along the HB. As indicated by the
arrows, the gaps are still present in the outer fields 
(cf. discussion in Sect.~\ref{hb}). Also the blue straggler (BS)
sequence is clearly defined. The BS sequence seems to be more extended
in magnitude in the inner field (Fig.~\ref{cmd_hst}) than in the outer
fields (Fig.~\ref{cmd_bv_dan}).  In the groundbased data there is a
strong contamination by field stars, as expected from the low galactic
latitude of NGC 2808 (Table~\ref{ph_tab}). As discussed by W99, there is also some
differential reddening in the direction of this cluster: in
Figs.~\ref{cmd_bv_dan}$-$\ref{cmd_vi}, and in the following, we did
not apply any differential reddening correction.
\subsection{Comparison with previous work\label{obsc}}
\begin{figure}
{\par\centering \resizebox*{1\columnwidth}{!}{\includegraphics{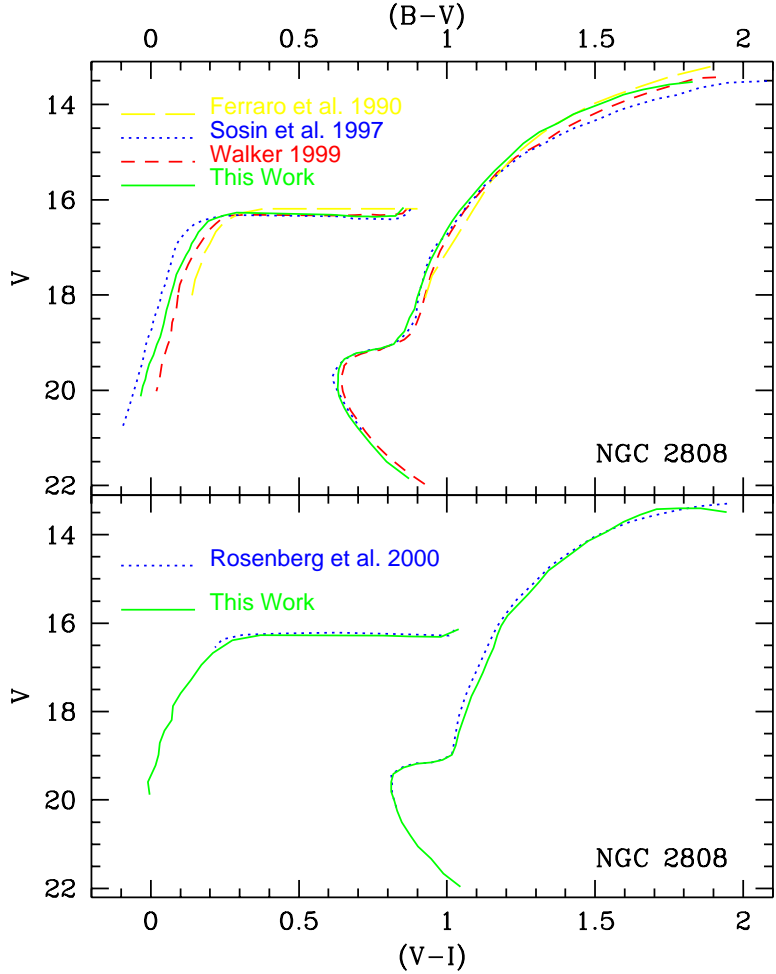}} \par}
\caption{In the {\it upper panel} the fiducial points from the CMDs of
Fig.~\ref{cmd_bv_dan} ({\it full lines}) are compared with the
fiducial points from the $V$ vs. ($B-V$) CMDs of S97 ({\it dotted
lines}), W99 ({\it dashed lines}) and F90 ({\it long dashed lines}).
In the {\it bottom panel} the comparison, on the $V$ vs. $V-I$ plane, of
the fiducial from Fig.~\ref{cmd_vi} ({\it full lines}) and Rosenberg et
al. \protect\cite*{rose00a} ({\it dotted lines}).
\label{fid_cmd}}
\end{figure}

The peculiar HB of NGC 2808 has called the attention of many
investigators in the past. The most complete CCD $V$ vs. ($B-V$) CMDs
have been published by Ferraro et al. \cite*{ferr90}(F90), S97,
W99. The hand drawn fiducial lines (on the original data) for the CMDs
of S97, W99, and Fig.~\ref{cmd_bv_dan}, and the fiducials given by F90
are compared in Fig.~\ref{fid_cmd} ({\it upper panel}). We must note
that the broadening of the sequences due to the differential reddening
makes the task of determining the fiducial lines quite difficult. An
uncertainty of the order of at least $\pm0.02$ magnitudes must be
associated with the fiducial points discussed below. First of all, we
note a general agreement with the photometry of W99 from the MS to the
RGB tip. A shift by 0.02 magnitudes (most likely due to differential
reddening) would make the two fiducial lines overlap, except only 
for the brightest part of the RGB and the faint part of
the HB, where it is quite hard to get accurate fiducials because of
the small number of stars.  In any case, a small color term error at
these extreme colors cannot be excluded.  In this regard, we note 
that while the color coverage of the standards used by W99 was
$-0.2<$($B-V$)$<1.5$, our calibration is based on a much larger number
of standards with $-0.3<$($B-V$)$<1.6$ (cf. Sect.~\ref{obsc}).  As
expected from the general agreement with W99, Fig.~\ref{fid_cmd}
confirms that the photometries of S97 and F90 are affected by both
scale and zero point errors (cf. W99 for more details). As already
discussed in the previous section, in view of the problems in
calibrating the $HST$ data of S97, in the present paper we have linked
the $HST$ magnitudes directly to the more robust groundbased system.
Fig.~\ref{fid_cmd} ({\it lower panel}) compares the fiducial lines of
Fig.~\ref{cmd_vi} with the CMD from Rosenberg et al.
\cite*{rose00a}. There is a general agreement between the two sets of 
photometry, with a possible small ($0.01\div0.02$ magnitudes) zero
point difference in color, which is likely due to differential
reddening.
\section{Reddening, distance, metallicity }
\label{obsR}
As discussed in W99, the reddening of NGC 2808 is very
uncertain. Surely, some of the discrepancies among the different
values in the literature are due to the effects of the differential
reddening. Also on the value of differential reddening there is no
general agreement.  We have estimated the differential reddening
in the region from the cluster center to $r=400$ arcsec by comparing
the dispersion of the RGB for $13<V<14$ and the dispersion due to
the photometric errors calculated using the artificial star tests.
The resulting differential reddening is 0.02 magnitudes. 

In Sect.~\ref{anRGB}, we will adopt a new and original method to get the
reddening of NGC 2808, following the analysis of the RGB morphology in
the $V$ vs. ($V-I$) plane presented in Saviane \& Rosenberg (1999) and 
Saviane et al. \cite*{savi00a}. In the following, we will make use of a
CMD obtained from the stars in an annulus centered on the cluster
center and covering a radial interval $30<r<200$ arcsec.  The
reddening that we will obtain must be considered as the average
reddening within this annulus.  As discussed below, for a proper use
of the method described by Saviane et al. \cite*{savi00a} (hereafter
S00) for the reddening estimate, we will have to assume a metallicity
for NGC 2808. To be consistent with S00, we will adopt the
metallicities given by Rutledge et al. 
\cite*{rutl97}(RHS97), i.e. [Fe/H]=$-1.24\pm0.03$ on the Carretta \& 
Gratton
\cite*{carr97}(CG) scale, as extended by Cohen et al
\cite*{cohe99}, and [Fe/H] $=$ $-1.36$ $\pm0.05$ on the
Zinn \& West \cite*{zinn84}(ZW) scale.
\subsection{Reddening and distance from the RGB stars}
\label{anRGB}
\begin{figure*}
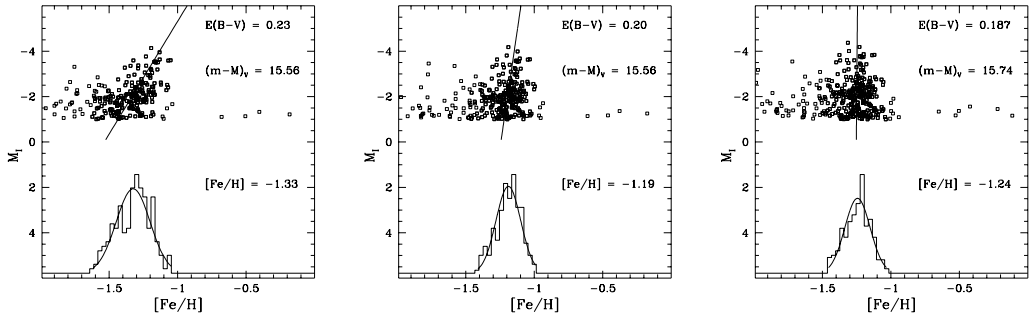

{\centering \begin{tabular}{ccc}
\resizebox*{0.32\textwidth}{!}{\includegraphics{H2150.f08}}  &
\resizebox*{0.32\textwidth}{!}{\includegraphics{H2150.f09}}  &
\resizebox*{0.32\textwidth}{!}{\includegraphics{H2150.f10}}  \\
\end{tabular}\par}
\caption{ [Fe/H] vs. $M_{I}$  for different
distance moduli and reddenings. In the lower part of the three panels
there is the histogram of the distribution in {[}Fe/H{]}. The labels
in each panel show the adopted reddening and distance, and the
resulting mean metallicity.  The Gaussian computed with the obtained
mean and $\sigma$ is overplotted to each histogram.
\label{red1}}
\end{figure*}

\begin{figure*}
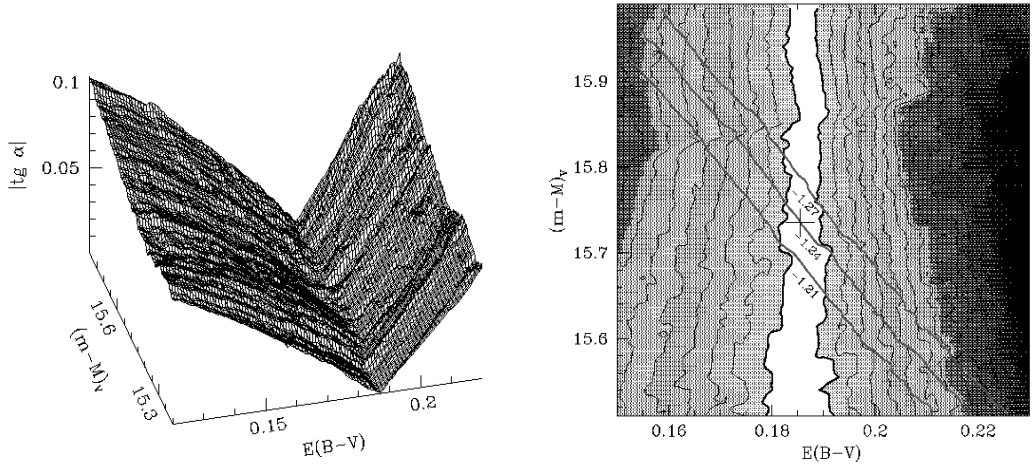

{\centering \begin{tabular}{cc}
\resizebox*{0.5\textwidth}{!}{\includegraphics{H2150.f11}} &
\resizebox*{0.5\textwidth}{!}{\includegraphics{H2150.f12}} \\
\end{tabular}\par}
\caption{The $|{\rm tg}\alpha |$ parameter (see text) is plotted as 
a function of the reddening and distance modulus as a 3d surface ({\it
left panel}), and contour map ({\it right panel}). The best reddening
is well defined by a valley of minimum values of $ |{\rm tg}\alpha|$,
centered at $E_{B-V}$ $\sim$0.185. Resolution of computed matrix is
0.001 and 0.005 in color and distance modulus, respectively.  The {\it
inclined heavy lines} on the contour map of the {\it right panel}
connect the points of the reddening-distance grid corresponding to a
mean metallicity (as calculated from Eq.~(\ref{equaz:radice1})) $[Fe/H]$
= $-1.24\pm0.03$. The almost {\it vertical heavy lines} show the 1
standard deviation from the minimum value of $ |{\rm tg}\alpha |$. The
``$+$'' marks the distance modulus and reddening used to produce the
{\it right } panel in Fig.~\ref{red1}.}
\label{tgalpha}
\end{figure*}

In S00 we used our photometric \emph{V, I} database of GGCs
(Rosenberg et al. 2000)
to define a grid of fiducial RGBs, which was then
used to find a monoparametric description of the red giant branches. The adopted
fitting function is $ M_{I}=a+b\cdot (V-I)+c/{[}(V-I)-d{]}$, 
where $ M_{I} $ and $ (V-I)$ 
are the absolute magnitude and dereddened color and 
each coefficient is a second-order polynomial in [Fe/H], i.e. 
$a=k1\, [{\rm Fe/H}]^{2}+k2\,[{\rm Fe/H}]+k3$, etc. (the coefficients are 
listed in Tab.~5 of S00). The calculations were repeated for two distance
scales and two metallicity scales (CG and ZW). The formula can now be inverted
in order to express [Fe/H] in terms of $ M_{I}$ 
and $ (V-I)$. Trivial algebra
yields: 
$$ \rm A[\rm Fe/H]^{2}+\rm B [\rm Fe/H]+\rm C=0, $$
where:\\

\(
\begin{array}{c} 
{\rm A}=k_{4}(V-I)^{2}+(k_{1}-k_{4}k_{10})(V-I)-k_{1}k_{10}\\
{\rm B}=k_{5}(V-I)^{2}+(k_{2}-k_{2}k_{10}-k_{5}k_{10})(V-I)\\
{\rm C}=k_{6}(V-I)^{2}+(k_{3}-k_{6}k_{10}-M_{I})(V-I)+\\
+k_{10}(M_{I}-k_{3}) 
\end{array} 
\) \\
Solving the quadratic equation, one finds
\begin{equation} 
\label{equaz:radice1} 
[{\rm Fe/H}]=(-{\rm B}+\sqrt{\Delta })/2{\rm A}.  
\end{equation} 
Where, as usually, ${\Delta ={\rm B}^{2}-4\, {\rm A}\, {\rm C}}$.  The
other root of the equation has no physical meaning. 
In this way, it is
straightforward to calculated the metallicity for any RGB star, provided
that the distance modulus and reddening are known. 
Once the correct combination of reddening and distance modulus are
used, in the [Fe/H] vs $M_{I}$ plane (Fig.~\ref{red1}), one expects a
vertical distribution, centered on the mean cluster metallicity, with
a dispersion which is related to the actual color dispersion in the
original CMD.
The {\it left panel} of Fig.~\ref{red1} shows the [Fe/H] vs $M_{I}$
diagram obtained from Eq.~(\ref{equaz:radice1}), based on an apparent
distance modulus $ (m-M)_{V}=15.56 $ and a reddening $E_{B-V}=0.23$
from the updated on-line catalog of Harris (June 1999 revision).  It
is evident that the RGB is not vertical. The situation improves
using the $E_{B-V}=0.20$ proposed by W99 (Fig.~\ref{red1}, {\it
central panel}), but still we need to search for a better combination
of parameters by exploring a possible range of reddenings and distance
moduli.
In order to have a quantitative measure of the best reddening and
distance modulus, we used the slope of the linear fit of RGB stars,  
$ |{\rm tg}\alpha| = {\rm \Delta{[Fe/H]} / \Delta M_{I} } $, 
in the absolute magnitude-metallicity plane. 
The dispersion $\sigma _{\rm {[Fe/H]}} $ in {{[}}Fe/H{{]}} around the
mean value, in the same plane, was also used; this parameter yields a
result consistent with that obtained from $|{\rm tg}\alpha |$, though
with a lower ``resolution''.
The values of $| {\rm tg}\alpha |$ were
computed for a discrete range of $ (m-M)_{V} $ 
and $ E_{B-V} $ (with steps of 0.001 in color and 0.005 in distance modulus). 
The $ E_{V-I} $  values were obtained using the relation 
$E_{V-I}=1.28\, E_{B-V}$  
\cite{dean78}. We started our calculations on the CG metallicity scale, and adopting
the Carretta et al. \cite*{carr99} distance scale (S00). On this
distance scale, the HB luminosity - metallicity relation is 
$M_{V}^{\rm HB}=0.18\, [\rm Fe/H]+0.90 $.
The value of $| {\rm tg}\alpha | $ as a function of the reddening and distance
modulus is plotted in Fig.~\ref{tgalpha} as a 3d surface ({\it left
panel}), and contour map ({\it right panel}). Clearly, the best
reddening is well defined by a valley of minimum values of $| {\rm
tg}\alpha | $, centered at $ E_{B-V}$ $\sim$0.185.  In order to
constrain also the distance modulus, we need to assume a metallicity.
In Fig.~\ref{tgalpha} ({\it right panel}), the {\it inclined heavy
lines} connect the points of the reddening-distance grid corresponding
to a mean metallicity (as calculated from Eq.~(\ref{equaz:radice1}))
[Fe/H]=$ -1.24$ (central line) $\pm 0.03$ (the two lateral lines).
The almost {\it vertical heavy lines} show the 1 standard deviation
from the minimum value of $|{\rm tg}\alpha| $, and can be used to
quantify the uncertainty on the reddening. Fig.~\ref{tgalpha} shows
that there is only a small allowed region for $(m-M)_{V} $ and $
E_{B-V} $, in order to have the mean metallicity from
Eq.~(\ref{equaz:radice1}) consistent with the adopted [Fe/H] value. 
In conclusion, a self consistent set of distance, reddening and
metallicity values is obtained only adopting $ E_{B-V}=0.19\pm0.01 $,
$ (m-M)_{V}=15.74\pm0.10 $, (internal errors) and assuming the RHS97
metallicity on the CG scale.  The error on the apparent distance
modulus was obtained as 1/2 the permitted range in Fig.~\ref{tgalpha}
({\it right panel}). The corresponding $ M_{I} $ vs. [Fe/H] diagram is
shown in Fig.~\ref{red1} ({\it right panel}).
The procedure was repeated for the ZW metallicity scale, obtaining a
slightly lower $ E_{B-V}=0.18\pm 0.01 $, and $ (m-M)_{V}=15.60\pm0.10$
(internal errors), still consistent with the previous determination
within the errors.
\subsection{Comparison with NGC 1851}
\begin{figure}
{\par\centering \resizebox*{1\columnwidth}{!}{\includegraphics{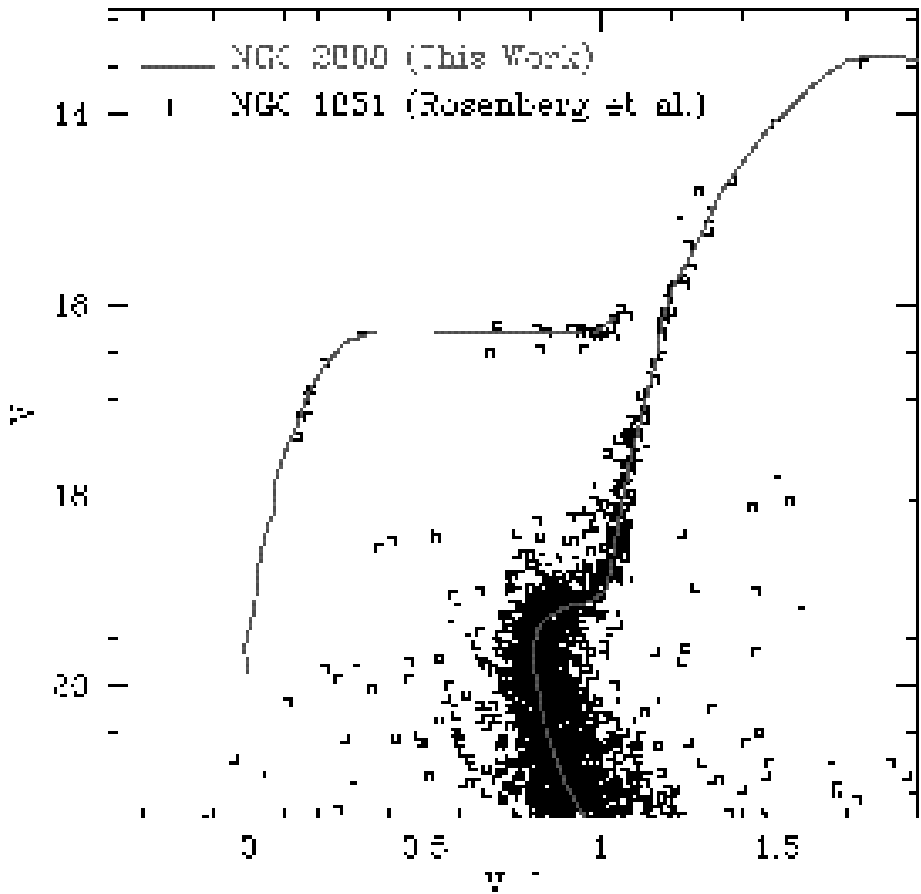}} \par}
\caption{The fiducial points from the CMD of Fig.~\ref{cmd_vi} ({\it full lines}) are
compared with the CMD of NGC 1851 from Rosenberg et al. \protect\cite*{rose00a}. A shift of
0.18 magnitudes in ($V-I$) and of 0.11 magnitudes in $V$ has been applied. 
\label{conf1851}}
\end{figure}

An independent check on the reddening and distance obtained above was
carried out by comparing the CMDs of NGC 2808 and NGC 1851. For NGC
1851, we used the CMD of Rosenberg et al. \cite*{rose00a}, whose
photometry is consistent with the present work, as shown in
Fig.~\ref{fid_cmd} ({\it lower panel}).  Moreover, according to RHS97,
the two clusters have the same metallicity within 0.08 dex (CG scale),
with NGC 1851 more metal rich than NGC 2808, and both clusters have a
bimodal HB. NGC 1851 has a low, fairly well determined reddening $
E_{B-V} $ = 0.02 
(Saviane et al. 1998). The NGC 2808 fiducials ({\it solid
lines}) and the CMD of NGC 1851 ({\it open squares}) are overplotted
in Fig.~\ref{conf1851}. The best match is obtained by applying a shift
of $ \Delta V=0.11 $ in magnitude, and $ \Delta(V-I)=0.18 $ in color
to the NGC 1851 diagram.  The corresponding shift in $ (B-V) $ would
be $\Delta E_{B-V} = 0.18/1.28 = 0.14 \pm 0.02$, 
which implies $E_{B-V}({\rm NGC~ 2808}) = 0.16 \pm 0.03$. 
This value being consistent with the previous
determination. 
To be consistent with the distance modulus determined in the previous
section, we have to adopt the same distance scale, i.e. the scale of
Carretta et al. \cite*{carr99}, based on the relation
$$M_{V}=0.18[Fe/H]+0.90$$ for the absolute magnitude of the
ZAHB. Adopting [Fe/H]$=-1.14$ for NGC 1851 we have $M_{V,{\rm
ZAHB}}^{1851}=0.69$.  Walker \cite*{walk98} finds $V_{\rm
ZAHB}^{1851}=16.20$, while Zoccali et al. \cite*{zocc99} give $V_{\rm
ZAHB}^{1851}=16.33$. Adopting a $\overline{V_{\rm ZAHB}^{1851}}=16.27$
for NGC 1851, we have an apparent distance modulus
$(m-M)_{V}^{1851}=15.58$ for this cluster. On the other side, NGC 1851
is 0.1 dex more metal rich than NGC 2808, and therefore it has a 0.02
magnitudes fainter HB. In summary, we have $(m-M)_{ V}^{2808}$ =
$(m-M)_{V}^{1851}$ $+$ ${\Delta}{M}_{V}$(ZAHB) $+$ ${\Delta}{V}$ =
$15.58 + 0.02 + 0.11$ = $15.71\pm0.20$, where the error accounts for
all the error sources.
A similar approach to estimate the reddening of NGC 2808 has been
followed by W99, who used as template RGB or HB sequences the CMDs of
three well-studied clusters (NGC 6362, NGC 1851, and NGC 6229) having
similar metallicities.  His best final value for the reddening is
$E_{B-V}=0.20\pm0.02$, but he also notices that the shape of the RGB
would suggest a slightly lower reddening and higher metallicity. We
now have been able to show that the RGB morphology does suggest a
lower reddening ($E_{B-V}=0.18\div 0.19 $, depending on the
metallicity scale), still compatible with the spectroscopic value of
[Fe/H].
\begin{figure}
{\par\centering \resizebox*{1\columnwidth}{!}{\includegraphics{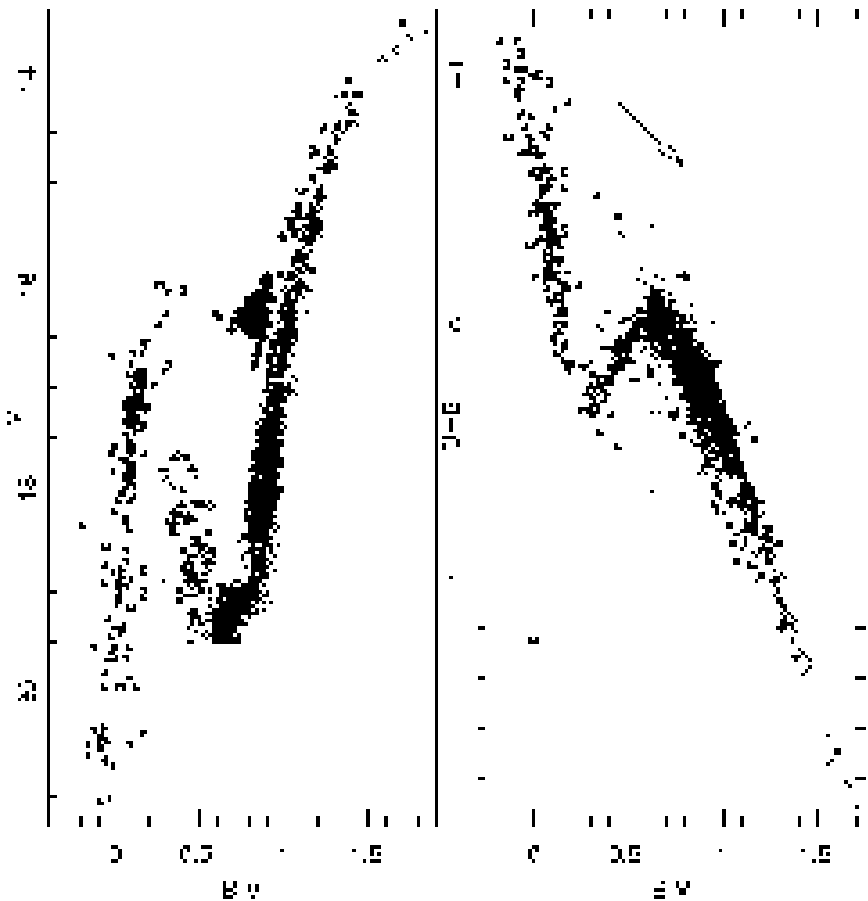}} \par}
\caption{ The $(V, B-V)$ CMD ({\it left panel}) and the
$(U-B, B-V)$ two color diagram for the $Danish$ stars ({\it right
panel}) between 100 and 400 arcsec and with a photometric error less
than 0.05 magnitudes in $(B-V)$ color.  The same symbols have been
used in the left and right panels in order to make easier to identify
the different evolving sequences in the two-color
diagram. \label{2col}}
\end{figure}
\subsection{Comparison with theoretical models}
The comparison between theoretical predictions and multiband 
photometric data collected for NGC 2808 allows us to supply independent 
estimates of both reddening and distance modulus. 
A similar approach has been used also by S97 who
present a detailed comparison between ZAHB models 
(Dorman et al. 1993) 
and $HST$ observations.  However, their main conclusions
concerning the reddening and the distance modulus were
hampered by the problems already mentioned in the zero-point of the
photometric calibration and in the transformation of theoretical
predictions into the observational plane.

First, we will start with the classical two-color diagram $(U-B, B-V)$
of Fig.~\ref{2col}, to constrain the cluster reddening, since it does
not depend on the distance modulus. In particular, we will focus our
attention on the HB stars, because their distribution in this plane
is strongly affected by reddening (Fig.~\ref{2col}).
\begin{figure}
{\par\centering \resizebox*{1\columnwidth}{!}{\includegraphics{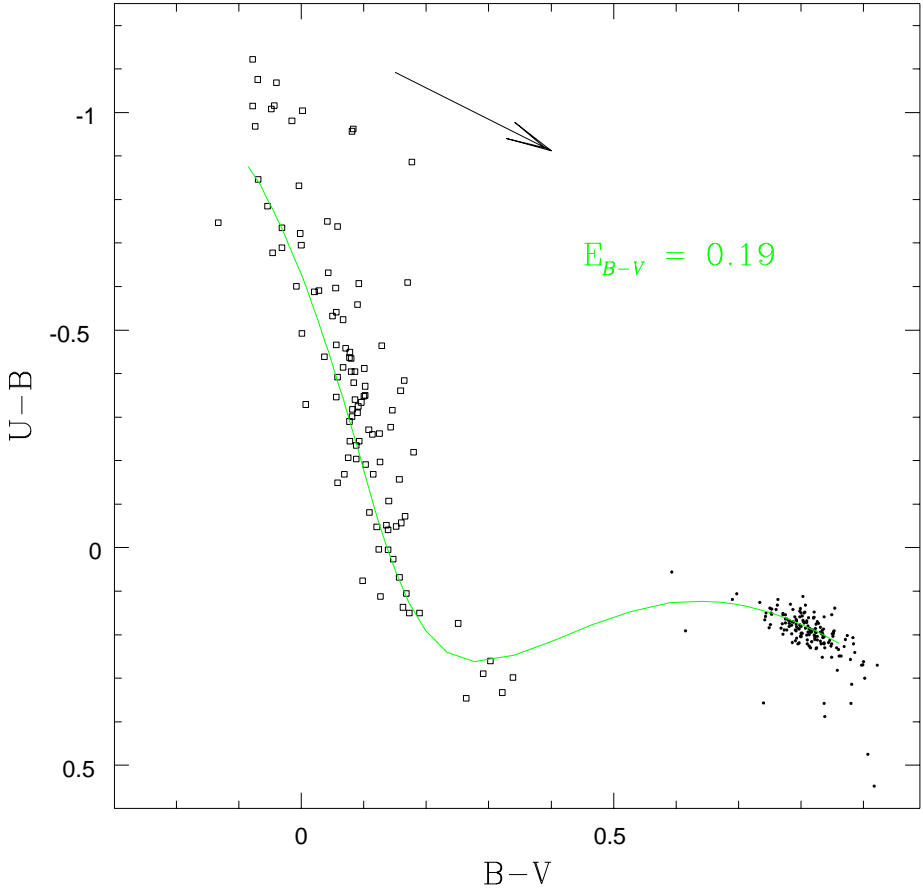}} \par}
\caption{Comparison between the observed HB stars ({\it open squares} for blue, and 
{\it dots} for red clump) and the models by Cassisi et
al. \protect\cite*{cass99} in the $(U-B, B-V)$ plane.  Only stars
between 100 and 400 arcsec with a photometric error less than 0.03
either in $(B-V)$ and $(U-B)$ magnitudes are plotted. The best fit is
obtained assuming an $E_{B-V}=0.19$.\label{X}}
\end{figure}

Fig.~\ref{X} shows the comparison in the $(U-B,B-V)$ plane between the
Danish sample and the theoretical prescriptions about the ZAHB locus
for Z=0.001 and an initial helium content equal to Y=0.23 (Cassisi et
al. 1999).
The ZAHB models correspond to a RGB progenitor with mass equal to
$0.8M_\odot$, whose He core mass at the He ignition and surface He
abundance after the first dredge up, are equal to $0.5018M_\odot$ and
$Y_{\rm HB}=0.243$, respectively.  The minimum stellar mass plotted in
Fig.~\ref{X} is equal to $0.5022M_\odot$; with an effective
temperature $\approx35,000$K lower than the one corresponding to the
He Main Sequence.

Bolometric magnitudes and effective
temperatures were transformed into the observational plane by adopting
the bolometric corrections and the color-temperature relations
provided by Castelli et al. \cite*{cast97a,cast97b}. 
By adopting the extinction curve by   
Cardelli et al. 
\cite*{card89}, 
we obtain a plausible fit (Fig.~\ref{X}) between theory and
observations only by assuming $E_{B-V}=0.19\pm0.03$ (solid line).
 
Once the reddening value is secured, we can estimate the distance 
modulus of NGC 2808 by comparing the theoretical ZAHB and the observed
HB stars in the $(V, B-V)$ CMD. 
\begin{figure}
{\par\centering \resizebox*{1\columnwidth}{!}{\includegraphics{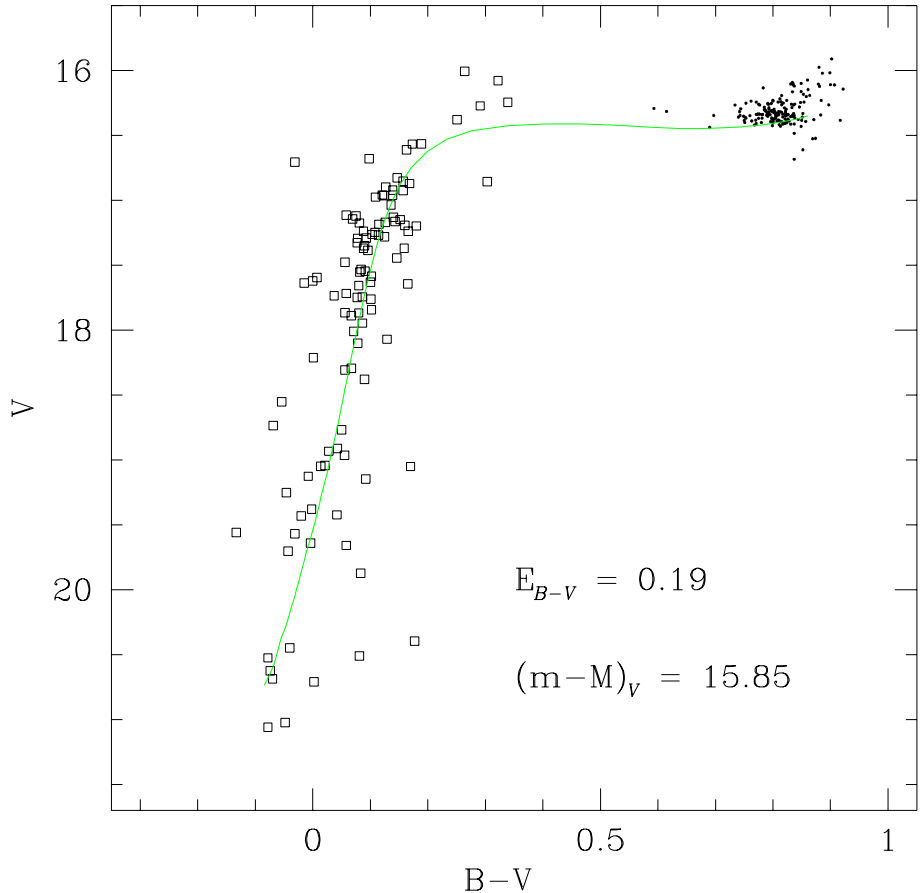}} \par}
\caption{Comparison between the observed HB (symbols as in Fig.~\ref{X}) 
and the models by Cassisi et al. (1999) in the $(V, B-V)$ plane. Stars
are selected as in previous figure.  By adopting a reddening
$E_{B-V}=0.19$ (cf. Fig.~\ref{X}), the best fitting apparent distance
modulus is ($m-M$)$_V=15.85$.\label{Y}}
\end{figure}
Fig. \ref{Y} shows that the models fit both the blue and red HB
stars if we assume a distance modulus equal to $15.85\pm0.1$ (fitting
errors only).  Both the reddening value and the distance modulus we
derived by comparing theory and observations are, within current
uncertainties, in fair agreement with the independent estimates we
obtained in the previous sections.

As a consequence, in the following we will adopt a reddening value of
$E_{B-V}=0.19\pm0.01\pm0.02$ (to take into account the differential
reddening), and an apparent $V$ distance modulus $(m-M)_V=15.79\pm0.07$
(internal error) $\pm0.08$ (due to the differential reddening).

\section{The Horizontal Branch morphology\label{hb}}
We have already discussed how the HB peculiarities possibly make NGC
2808 a unique test case for understanding the formation and evolution
of the HB stars.  In this section, we will make use of our large 
sample to better define the stellar distribution along the HB 
and the spatial distribution of HB stars characterized by different 
effective tempertatures.   

All the CMDs of Figs.~\ref{cmd_bv_dan}$-$\ref{cmd_vi} clearly show
the EBT. In general, as a consequence of the photometric errors, the
EBT tends to become more and more dispersed toward fainter magnitudes,
though some intrinsic dispersion (at a level of $\sim0.02$ magnitudes
(after the subtraction of the contribution by the photometric error and
the differential reddening)) cannot be excluded.
By using our evolutionary models for He-burning structures we have
verified that the spread present in
Fig.~\ref{Y}, is mainly due to evolutionary effects.

Both the CMD for the inner 100 arcsec (Fig.~\ref{cmd_hst}) and those
for the outer regions (e.g.  Fig.~\ref{cmd_bv_dan}) show that the EBT
extends to $V$=21.2, confirming the results of W99. The EBT extends by
more than one magnitude in $U-B$ when moving from low to
high-temperatures.  The HB presents two well-defined gaps along the
blue tail, and even though a sizable fraction of HB stars lie on both
the red and blue side of the instability strip, the region in between
is poorly populated, and indeed only two RR Lyrae have been currently
identified (Clement
\& Hazen 1989).  Interestingly enough, the three gaps are also present
in both in the inner and outer regions, showing that they are not
confined to the inner core. 
This empirical evidence is worth being investigated in more detail, 
because it could be crucial to finding out the origin of these features.  
\subsection{The HB $U$-jump \label{jump}}
In the previous section, we have already presented a comparison 
between the observed HB and the models by Cassisi et al. \cite*{cass99}
in the $(U-B,B-V)$ and $(V, B-V)$ planes for an independent reddening
and distance determination. 
Note that theoretical predictions properly account for the ZA\-HB 
distribution of both cool and hot HB stars (see Fig.~\ref{X} and Fig.~\ref{Y}).  
%
\begin{figure}
{\par\centering \resizebox*{1\columnwidth}{!}{\includegraphics{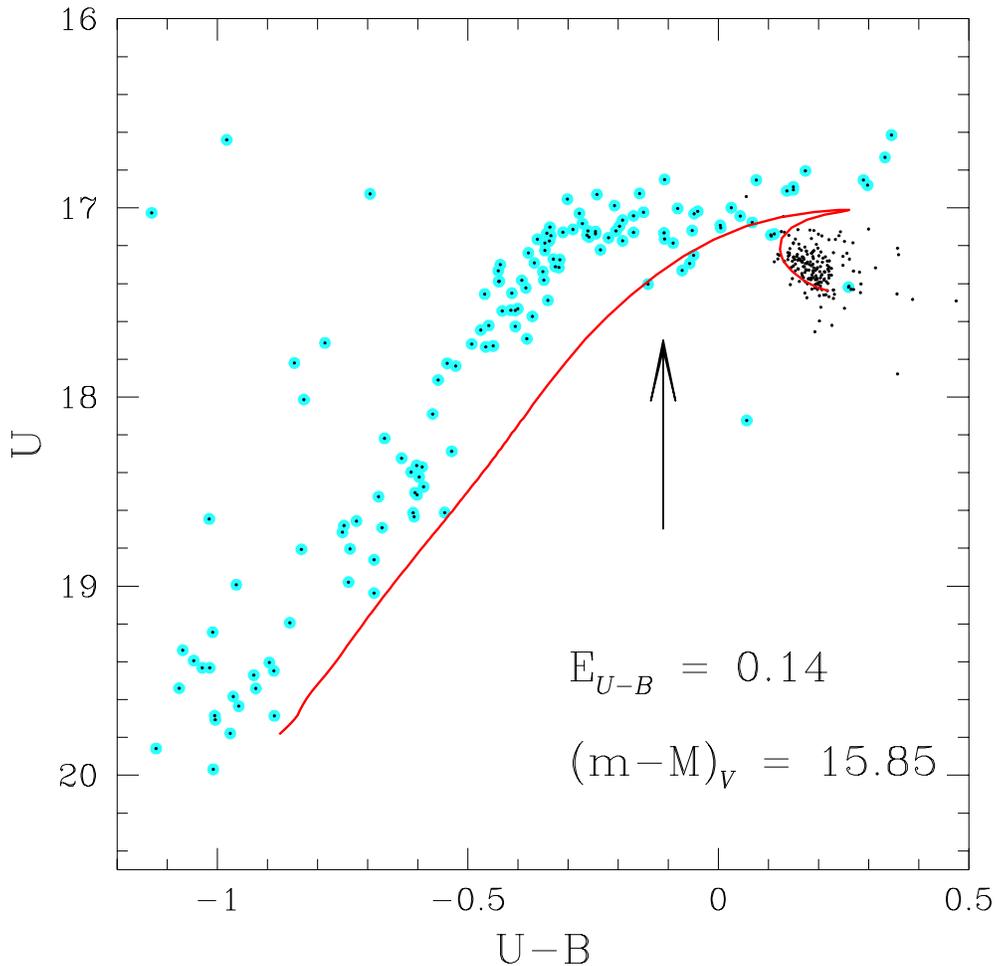}} \par}
\caption{Comparison between the observed HB ({\it filled circles} for blue 
and {\it dots} for red clump) stars and the models by Cassisi et
al. \protect\cite*{cass99} in the $(U, U-B)$ plane. Note that the jump
discovered by Grundhal et al. (1999) in the Str\"omgren $u$ band is
also visible in the Johnson $U$ band at $U-B \approx -0.11$ (indicated
by the arrow).  According to our HB models the jump is located at
$T_{\rm e}\sim11,600$.
\label{Z}}
\end{figure}
On the basis of accurate Str\"omgren photometric data, 
Grundahl 
\cite*{grun98}
and Grundahl et al. \cite*{grun99}
found that all GGCs with a sizable fraction of hot HB stars present a
well-defined jump in the $u$ band among blue HB stars.  In particular,
they succeeded in demonstrating that in the $(u, u-y)$ plane HB stars
characterized by effective temperatures ranging from 11,500 to 20,000
K are brighter and/or hotter than ca\-no\-ni\-cal HB models. This
effect was explained as an increase in the abundance of elements
heavier than carbon and nitrogen caused by radiation levitation. The
overall scenario was soundly confirmed by high-resolution
spectroscopic measurements
(Behr et al. 2000a; Behr et al. 2000b). 
Owing to the good accuracy of our ground based photometric data and
the large sample of HB stars we decided to test whether a similar
effect is also present in the ca\-no\-ni\-cal $(U,U-B)$ plane.
Fig. \ref{Z} shows quite clearly the appearance of a $U$-jump among
HB stars hotter than $U-B = -0.11\pm0.03$, thus suggesting that this
effect can also be detected in the ca\-no\-ni\-cal $U$ band.  This
finding is not surprising, since a detailed check of Fig. 7 in
Grundahl et al. \cite*{grun99} shows that the emergent flux in the $U$ and in the $u$
bands are quite similar for $11,500 \le T_{\rm e} \le 20,000$ K.
If we assume that the jump takes place at $U-B = -0.11\pm0.03$ 
the HB stars start to be affected by radiation levitation 
at effective temperatures of the order of $11,600\pm350$ K, in very
good agreement with the results by Grundahl et al. (1999).
This also confirms the spectroscopic results by Behr et al. (2000a,
and reference therein) suggesting that the bulk of the HB stars hotter
than 10,000 K present strong heavy element enhancements.
Even though color-temperature transformations are still affected by
systematic uncertainties in the short wavelength region, this finding
brings out the evidence that the effects of radiation levitation and
elemental diffusion in blue HB stars can also be investigated in the
$U$ band.
\subsection{Statistical significance of the gaps\label{signif}}
\begin{figure}
{\par\centering \resizebox*{1\columnwidth}{!}{\includegraphics{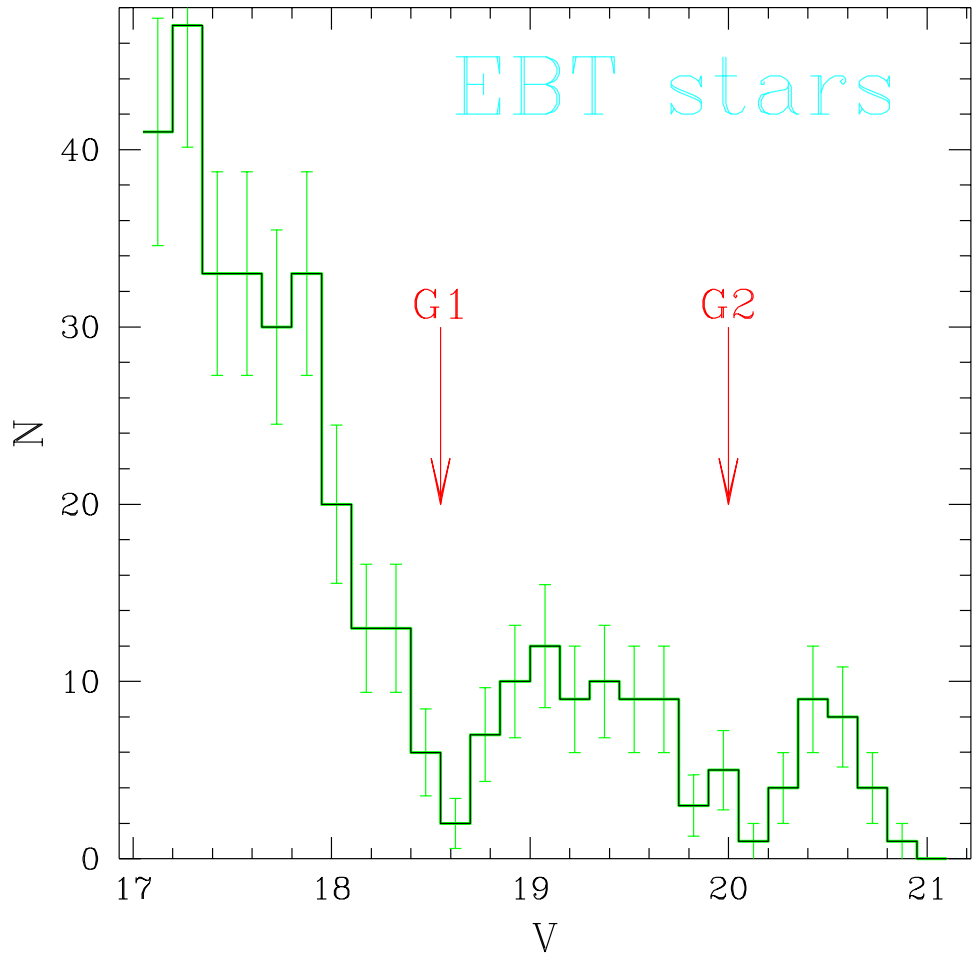}} \par}
\caption{Histogram for the 497 stars on the EBT 
with $17<V<21$ and a photometric error $<$ 0.05.  These stars have
been extracted from the CMDs of Fig.~\ref{cmd_hst} ($r<100$ arcsec)
and Fig.~\ref{cmd_bv_dan} ($100<r<1040$ arcsec).  The two arrows
indicate the position of gaps labelled with G1 and G2 (see text). The
bin size is 0.15 magnitudes in $V$.
\label{histo_hb}}
\end{figure}

As discussed in Catelan et al. \cite*{cate98}, assessing the
statistical significance of the gaps along the EBTs is not a trivial
task. Following the discussion in Piotto et al. \cite*{piot99a},
taking advantage of the fact that the EBT runs almost parallel to the
$V$-magnitude axes, we have constructed a histogram of the EBT star
distribution in $V$-magnitude. Fig.~\ref{histo_hb} shows this
histogram for the 497 stars on the EBT with $17<V<21$ and a
photometric error $<$ 0.05.  These stars have been extracted from the
CMDs of Fig.~\ref{cmd_hst} ($r<100$ arcsec) and Fig.~\ref{cmd_bv_dan}
($100<r<1040$ arcsec). Confirming the visual impression from the CMDs,
Fig.~\ref{histo_hb} shows a gap at $V\sim18.5$ (hereafter G1) and a
second one (G2) at $V\sim20.0$.  From a comparison with the models
(Fig.~\ref{Y}), G1 is located on the HB in a position corresponding to
a $T_{\rm e}\sim15,900 K$, while G2 is at $T_{\rm e}\sim25,400 K$. The
corresponding masses are 0.57 $M_{\odot}$ and 0.52 $M_{\odot}$
respectively.
Judging from the error bars (Poisson errors) of the histogram, we can
confirm the results of S97 on the significance of these gaps.
We can also follow a different approach (Catelan et al. 1998).
In the simplest hypothesis, assuming a uniform distribution in $V$ of
the stars along the EBT, we can calculate the probability $P_{\rm g}$
of having a gap at a given position.  Using Eq. 2 of Catelan et
al. (1998),
we find: $P_{\rm g}$=$3\times10^{-5}$ for G1 and $P_{\rm
g}$=$4\times10^{-6}$ for G2. This is the probability of having a gap
exactly at the position of G1 and G2 from a uniform
distribution. Indeed, we are interested to see what is the probability
to have a gap on the EBT, not a gap at a given position on the
EBT. Using Eq. (3) in Catelan et al. (1998) we have that the
probability $P_{\rm r}$ of finding a gap at any position on the HB of
Fig.~\ref{histo_hb} is still $P_{\rm r}=3\times10^{-3}$ for G1 and
$P_{\rm r}=3\times10^{-4}$ for G2.

In conclusion, if the distribution in $V$ of the stars along the EBT
is uniform, the probability of finding even only one of the gaps G1
and G2 is negligible.  This would imply that these gaps must be real
in some physical sense, and not a statistical fluctuation. However, we
do not know if the null hypothesis of a uniform distribution is
correct. Indeed, looking at the distribution of the stars in
Fig.~\ref{histo_hb}, and in the other GGCs with EBTs (Fig.~9 in Piotto
et al. 1999, see also Catelan et al. 1998), it seems that the stellar
distribution might be not uniform at all. Still, the fact that all the
clusters with an EBT show gaps along it, and that some of these gaps
seem to be located at very similar positions (either in temperature or
mass) on the HBs of different clusters (Piotto et al. 1999), might be
suggestive of a genuine physical mechanism for producing gaps.
A physical origin is furtherly strenghtened by the empirical results
by Behr et al. (2000a) and Behr et al. (2000b),
who find a discontinuity both in the rotational velocity and in the
abundance ratios of the stars at the level of the gap in the EBT of
M13. Though there have been many attempts to explain the observed gaps
(Catelan et al. 1998; Caloi 1999; D'Cruz et al. 1996; Rood et
al. 1998; Soker 1998),
we still lack a coherent and quantitative model which accounts for
their origin.  
\subsection{Radial distribution of the HB stars\label{radial}}
\begin{figure}
{\par\centering \resizebox*{1\columnwidth}{!}{\includegraphics{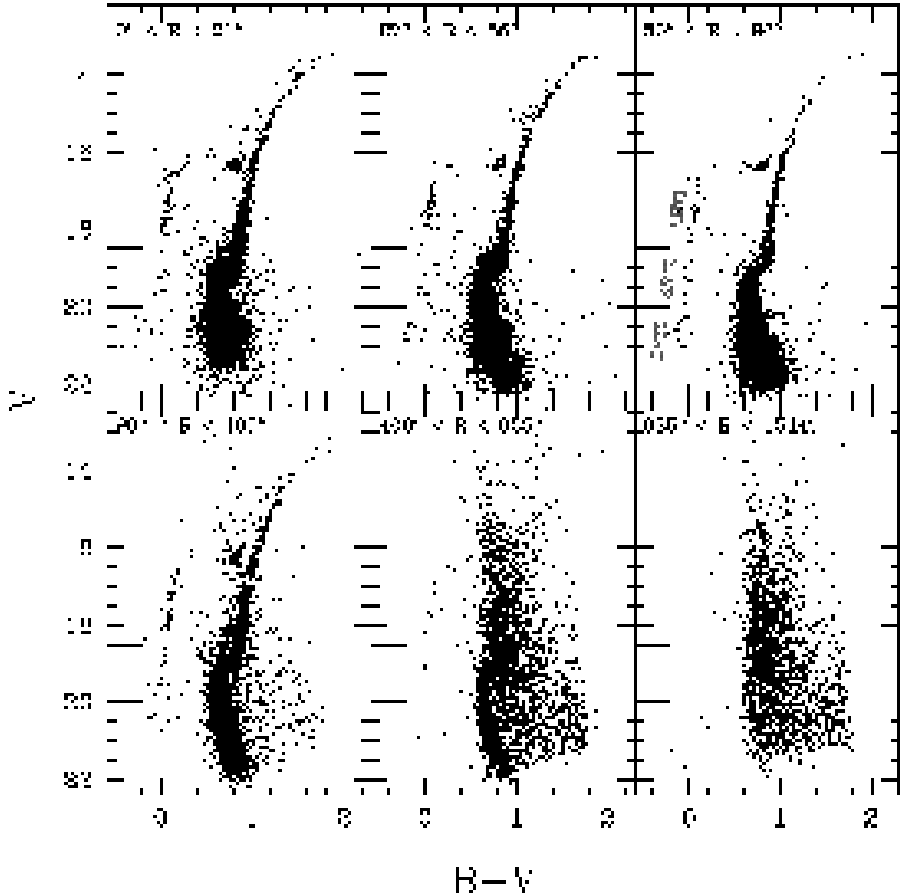}} \par}
\caption{$V$ vs. ($B-V$) CMDs in 6 radial bins. The CMDs in the {\it 
upper panels} come from the $HST$ data. The CMD in the {\it lower
right} panel extends from the tidal radius $r_{\rm t}$ to about
$1.7r_{\rm t}$. The EBT is present in all the five internal bins ($r<r_{\rm t}$),
and in all cases it extends down to $V=21.2$. Also the gaps seems to
be present everywhere inside the cluster, and at the same location on
the HB. The {\it upper right} panel shows also the location of the
three subgroups in which we have divided the EBT for the calculation
of the radial gradients (see text). \label{cmd_rad}}
\end{figure}

\begin{figure}
{\par\centering \resizebox*{1\columnwidth}{!}{\includegraphics{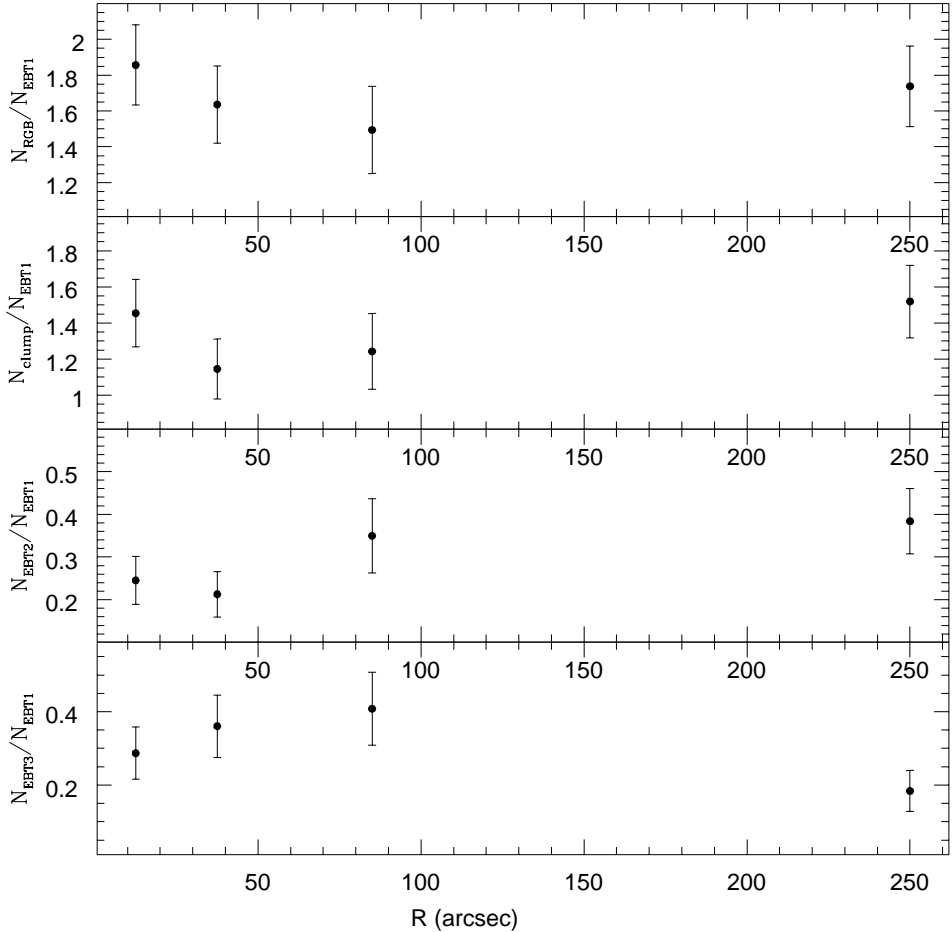}} \par}
\caption{Radial trend of the ratio of the RGB and HB stars
(corrected for complitness). Though no statistically significant gradient
is visible, we note the possible decrease of RGB stars out to 100
arcsec ({\it upper panel}), and the passible lack of the hottest HB
stars in the outer bin ({\it bottom panel}).
\label{grad}}
\end{figure}

Fig.~\ref{cmd_rad} shows the $V$ vs. ($B-V$) CMDs of NGC 2808 in six
different radial bins. The contamination by field stars becomes rather
strong in the outer fields, as expected from the location of the
cluster in the Galaxy. According to Trager et al. \cite*{trag93}, the
tidal radius of NGC 2808 is $r_{\rm t}=15'.6$. Therefore, we expect
all the stars in the bottom right panel of Fig.~\ref{cmd_rad} to be
field stars.
The most interesting facts shown by Fig.~\ref{cmd_rad} are that:
\begin{enumerate}
\item the EBT is present surely beyond $400$ arcsec, corresponding to more than
nine times the half mass radius $r_{\rm h}$, and it extends to $V=21.2$ also in
these external regions;
\item also the gaps on the EBT are present at least out to 400 arcsec from the
cluster center, and possibly beyond it;
\item the locations of the gaps in the CMD seem to be the same all over the cluster.
\end{enumerate}
We have also checked the relative number of HB stars as a function of
the radius. The HB stars have been divided into 4 groups (cf. 
Fig.~\ref{cmd_rad}): red clump (all the HB stars redder than the RR
Lyrae instability strip); EBT1 (the HB stars bluer than the RR Lyrae
instability strip, but brighter than the first gap G1); EBT2 (the HB
stars within the two gaps); EBT3 (the HB stars fainter than the gap
G2).
The relative numbers of these stars (corrected for complitness) out to
400 arcsec are shown in Fig.~\ref{grad}.  Overall, there is no
statistically significant radial gradient. This is the most important
observational evidence.

It must be noted that 
D'Cruz et al. (2000) and Whitney et al. (1998) found an absence of
radial gradients in the distribution of the EBT stars in $\omega$ Cen,
though in that case the investigation was extended only out to 0.3
tidal radii.  We want to furtherly comment on two of the panels of
Fig.~\ref{grad}. First of all, the {\it bottom panel} shows that the
faintest (hottest) HB stars (group EBT3) seems to be less numerous in
the outer bin ($r>100$ arcsec). However, the small number of stars and
the uncertainty on the completeness correction make this difference
very marginal (at 2 sigma level). Second, the upper panel seems to
show a marginal trend of the RGB\footnote{We consider RGB stars
brighter than HB red clump, assumed $V_{\rm clump}^{2808}=16.3$. }
stars (a decreasing of RGB stars with respect to the EBT1 ones, moving
outwards, up to $\sim100$ arcsec), which is opposite to what has been
found in the post core collapse\footnote{ Note, however, that NGC 2808
is a normal King model cluster.  }  clusters M30 by Piotto et
al. \cite*{piot88}, as recently confirmed also by Howell \&
Guhathakurta \cite*{howe00}, and M15 by Stetson \cite*{stet94}.
However, a possible population gradient like in the upper panel of
Fig.~\ref{grad} could explain the much more significant color gradient
found by Sohn et al. \cite*{sohn98} in NGC 2808, which goes in a sense
of a redder center.  Despite the fact that none of the gradients in
Fig.~\ref{grad} is statistically significant, they might still be real
in some physical sense.  Indeed, as throughly discussed by Djorgovski
\& Piotto \cite*{djor93a} the color and population gradient issue in GGCs is
always hampered by small number statistics and errors in the
completeness corrections. What can make them significant is the
occurence of similar gradients in a number of clusters. Before drawing
any conclusion, we need to extend this investigation to other clusters
with EBTs.

It has been suggested, and supported with observational evidences,
that the HB blue tails and the EBTs might be of dynamical origin. For
example, Fusi Pecci et al. \cite*{fusi93} have found a correlation between the HB
elongation and cluster central densities. Tidal stripping of the
envelopes of the stars evolving along the RGB during close encounters
in a high density environment can be a possibility
(Djorgovski \& Piotto 1993). 
Orbital angular momentum transfer to a RGB envelope in a close flyby
which can favour enhanced mass loss or even deep mixing (Sweigart
1997)
is another possibility. The difficulties in dealing with these
hydrodynamical phenomena has prevented till now a modeling of the
proposed mechanisms. Still, it is somehow surprising to see that EBT
stars can be found in NGC 2808 at $r>9r_{\rm h}$ (middle bottom panel
of Fig.~\ref{cmd_rad}), where the stellar densities are extremely low.
Of course, we cannot completely exclude the possibility that the EBT
stars have very elongated orbits, which bring them very close to the
cluster center.  However, the hypothesis of an high anisotropy for the
GC stellar orbits is still to be proved.
Another possibility is that the same mechanism responsible for the
envelope stripping can cause a kick off from the cluster core of some
of the involved stars. In any case, the absence of any clear gradient
in the radial distribution of the EBT stars can again be a major
obstacle for this mechanism. Alternatively, since most of the close
encounters will take place within the dense core of the cluster, one
could expect the EBT stars to be more centrally concentrated than the
red HB stars. This is not confirmed by Fig.~\ref{grad}, with a
possible exception of the extreme EBT3 stars, but only for $r>100$
arcsec, i.e. $r>2.2\,r_{\rm h}$.  Of course, it is possible that
dynamical relaxation causes the encounter products to quickly diffuse
out, smoothing their radial distribution. However, we find EBT stars
well beyond the half mass radius, while already the half mass
relaxation time ($t_{\rm h}=1.3\times10^{9}$ yr, Tab.~\ref{ph_tab}) is
sensibly longer than the evolutionary time on the HB ($10^8$ yr).  The
absence of gradients exclude also the possibility that EBT stars are
formed from tidal stripping in close binaries
(Bailyn \& Pinsonneault 1995), as, also in this case, we would expect
to see more EBT stars in the cluster core, where the more massive
binaries would concentrate, than in the outer envelope.

In conclusion, we still lack an explanation of the EBT in NGC 2808,
and other clusters. The observational facts presented in this paper
seem to make rather unlikely a possible dynamical origin of the
phenomenon. And it is not clear whether the presence of EBTs and gaps
on them are the manifestation of the same physical phenomenon.
\section{The Luminosity Function}
As clearly shown by Fig.~\ref{cmd_rad}, the contamination by field
stars obviously becomes stronger and stronger for increasing $r$: the
inner CMDs from the $HST$ data are virtually not affected by this
problem, which becomes severe for $r>100$ arcsec. In order to obtain a
LF from the groundbased data, we need to statistically subtract the
field stars.  As our coverage of NGC 2808 extends far beyond the tidal
radius, in principle this is not a problem, though we must not forget
that the presence of the differential reddening will limit the
accuracy of any attempt to statistically subtract field stars.  The
subtraction of the field objects has been performed on the $HST$ CMD,
and on the $Danish$ $V$ vs. ($B-V$) CMD in the region $100<r<400$
arcsec. For $r>400$, the small number of cluster stars and the large
number of field objects makes the data of no use for the LF.
\begin{figure}
{\par\centering \resizebox*{1\columnwidth}{!}{\includegraphics{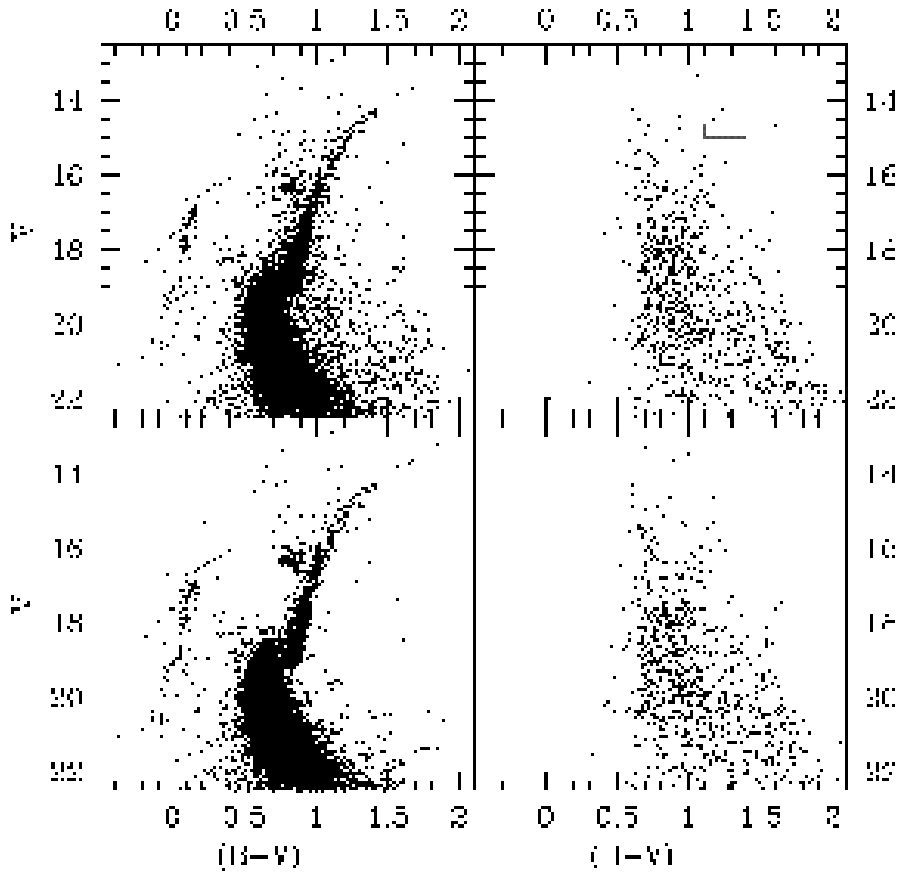}} \par}
\caption{The {\it upper-left panel} shows the original CMD from the
$Danish$ data with $100<r<400$. The {\it upper-right panel} shows the
template field star CMD, which refers to an area located on the
North-West side of the cluster, at $r>1.13\,r_{\rm t}$, and of the
same size of the area covered by the stars in the {\it upper-left
panel}. The field star subtracted CMD is in the {\it bottom-left
panel}. The {\it bottom-right panel} shows the CMD of the subtracted
stars. The small box in the {\it upper right} and in the {\it lower
left} panels shows the region of the heap (see text).
\label{sub_field}}
\end{figure}
For the $HST$ data, the number of subtracted stars is negligible (we
subtracted 74 stars out of 35,000 objects).

The results of the field subtraction on the $Danish$ data are
illustrated in Fig.~\ref{sub_field}.  The original CMD is plotted in
the {\it upper-left panel}. The {\it upper-right panel} shows the
template field star CMD, which refers to an area located on the
North-West side of the cluster, at $r>1.13\,r_{\rm t}$, and of the
same size as the area covered by the stars in the {\it upper-left
panel}. In the cluster CMD, for each field star we removed the star
closest in magnitude and color. The resulting field-subtracted cluster
CMD, and the CMD with the subtracted stars only are shown in the two
{\it bottom panels}.

\begin{figure*}
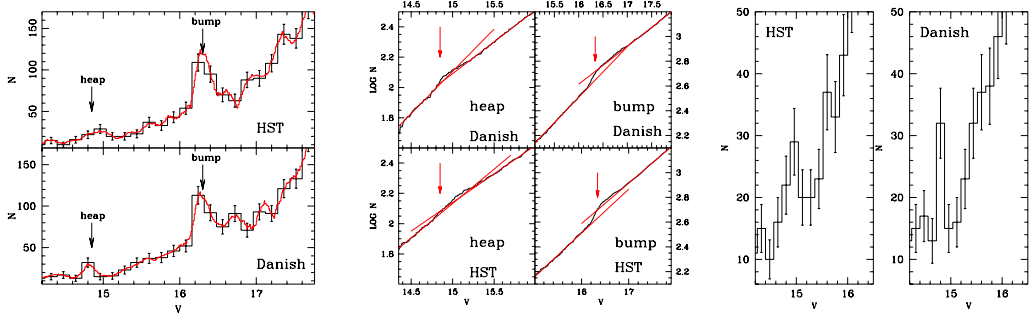

{\centering \begin{tabular}{ccc}
\resizebox*{0.32\textwidth}{!}{\includegraphics{H2150.f22}}      &
\resizebox*{0.32\textwidth}{!}{\includegraphics{H2150.f23}}      &
\resizebox*{0.32\textwidth}{!}{\includegraphics{H2150.f24}}  	 \\
\end{tabular}\par}
\caption{Differential ({\it left panel}) and cumulative 
({\it central panel}) LF, of the RGB of NGC 2808. There are two
features: the well known RGB bump and a newly discovered {\em heap}
(see the text).  
{\it Right panel} is an enlargement of {\it left panel}, with a different 
bin size, around the heap. \label{lf}}
\end{figure*}

The LF has been obtained from the field-subtracted CMD, excluding the
contribution of the blue stragglers, HB, and AGB stars.  A detailed
discussion of the LF in a large sample of globular cluster, including
NGC 2808, from the tip of the RGB to a few magnitudes below the TO,
and a comparison with the models, has been presented in Zoccali \&
Piotto (2000).  Here we focus our attention on the brightest part of
the LF (Fig.~\ref{lf}, {\em left panel}).  In particular, we note that
on both the $HST$ and $Danish$ LFs there are two features: the well
defined RGB bump, at $V\sim16.3$, and at $\sim1.4$ magnitudes brighter
a second bump (that we will call {\it heap} hereafter), at
V$\sim$14.9.  The heap is more clearly visible in the $HST$ data,
where the contamination by field stars is negligible.

The bump is a well known feature, due to the hydrogen burning shell
approaching the composition discontinuity left by the deepest
penetration of the convective envelope. Bump properties are thoroughly
discussed in Zoccali et al. \cite*{zocc99} and Zoccali \& Piotto
\cite*{zocc00}.
On the basis of the differential LF it is not possible to assess
whether the heap is a real feature or more simply a statistical
fluctuation. Indeed, also at magnitudes fainter than the bump the LF
presents other secondary peaks.

In order to shed more light on this problem, in Fig.~\ref{lf} we show
also the cumulative logarithmic LF ({\it mid\-dle pa\-nel}), since a
local increase in the number of stars causes a change in its slope.
The two slopes which fit the integral LF at magnitudes
fainter/brighter than $V \approx 14.9$ further support the evidence
that there is a real feature.
Note that the heap is also present in the LF in the $B$ band, at
$B\approx16.2$.
We are not aware of any previous detection of this heap. However, an
indication of the presence of the heap is also present in the data by
W99.

The physical nature of the heap can not be firmly constrained on the
basis of the present data alone. This notwithstanding, in the
following we supply some hints which can help to disclose the physical
mechanisms which might cause it.  First of all, we exclude the
interpretation that heap is a consequence of the contamination by
field stars, since it is located in a region of the CMD which is
marginally affected by this problem (cf. small box in the {\it
upper-left panel} of Fig.~\ref{sub_field}).  Similarly, it cannot be
due to contamination by AGB stars, since, at this magnitude, AGB and
RGB stars are well separated, and also because the expected ratio
between AGB and RGB stars at Z=0.001 is of the order of $0.15-0.20$.

The heap could be explained by assuming that it is the aftermath of a
deep-mixing episode which somehow causes a temporary delay in the
advancement of the H-shell. Even though this phenomenon has been
suggested for explaining peculiar metallicity overabundances in
several cluster RGB stars (Gratton et al. 2000)
and some intrinsic features of the extreme HB stars (Sweigart 1997;
Sweigart \& Catelan 1998)
the physical mechanisms which trigger its appearance are still
questioned and widely debated in the current literature (Grundahl et
al. 1999, and references therein).
However, taking into account the number of stars in the heap region
(in the $HST$ data, where the contamination by field stars is
negligible), if we assume as working hypothesis that the heap is
caused by a deep mixing episode, then its occurrence would imply an
increase in the evolutionary time during this phase of the order of
25\% when compared with the canonical one.

As a consequence, we decided to test whether the heap is a unique
feature of NGC 2808 or is present in other GGCs. We selected 47 Tuc,
since for this cluster we can use a homogeneous set of $HST$ data
reduced by the same software, and find heap at the same position above
the bump ($\Delta V \sim 1.4$),
and it is located, within the uncertainties, in the same region of the
47 Tuc RGB in which Edmonds \& Gilliland \cite*{edmo96}, on the basis
of $HST$ time series data, discovered for the first time K giant
variables (KGVs) in a GGC. In fact, the clumping of KGVs they found is
located at $M_V\sim-0.9$, which is in remarkable agreement with the
heap absolute magnitude, i.e. $M_V\sim-1.0$. The same outcome applies
to the $B-V$ color, and indeed the bulk of KGVs in 47 Tuc are located
at $B-V$ colors ranging from 1.1 to 1.2 (see Fig. 1 in Edmonds \&
Gilliland), which are quite similar to the color range covered by the
heap. Unfortunately, our data do not allow us to identify any
variability, since KGVs are characterized by low luminosity
amplitudes, and by short periods \cite{edmo99}.

It is also worth noting that the heap stars, as the KGVs, are not
centrally concentrated, and indeed the heap appears both in the
$Danish$ and in the $HST$ sample. If the hypothesis of a link between
the heap and the KGVs is confirmed by detailed time series multiband
photometric data, it could have a fundamental impact on our
understanding of the physical nature of these intriguing objects. In
fact, the appearance of the heap seems to suggest that KGVs spend a
substantial portion of the pulsation cycle when they are cooler.
Note that the change in the effective temperature to explain this
effect should be approximately 100-150 K at the typical colors of heap
stars. Moreover and even more importantly, if the heap is caused by
KGVs, the results on NGC 2808 show that the physical mechanisms which
trigger the pulsation destabilization in these objects are also
present at intermediate metallicities.
Finally, we mention that radial velocity variations have been detected
and measured in several field, intermediate-mass K giant stars (Hatzes
\& Cochran \cite*{hatz99}, and references therein).  These objects
often present mixed-mode behavior with periods ranging from a fraction
of day ($\beta$ Ophiuchi, Hatzes \& Cochran \cite*{hatz94a}) to few days
($\alpha$ Bootis, Hatzes \& Cochran \cite*{hatz94b}) or hundreds days
($\pi$ Herculis, Hatzes \& Cochran \cite*{hatz99}) and radial velocity
amplitudes ranging from few tens of $m/s$ to hundreds of $m/s$.  This
behavior was soundly confirmed by Buzasi et al. \cite*{buza00} who detected
and measured multimode oscillations in $\alpha$ Ursae Majoris with a
fundamental period of 6.35 days and an infrared luminosity amplitude
of 390 $\mu$mag. 

The lack of a comprehensive pulsation scenario which accounts for the
empirical behavior of these objects, and of firm constraints on the
occurrence of an instability strip prompts for new theoretical and
empirical investigations.  In this context, it is noteworthy that in a
recent investigation based on both Hipparcos and OGLE data, Koen \&
Laney \cite*{koen00} found evidence that pulsation in very high
overtones seems to be the most plausible mechanism to explain the
luminosity variation in field M giant stars with periods shorter than
10 days.
\begin{acknowledgements}
We thank the referee, Dr. N. D'Cruz, for the careful reading of the
manuscript and for the useful comments and suggestions which surely
helped to improve the paper.  We thank Dr. A. Walker for providing his
data on NGC 2808 in a computer readable form.  We acknowledge the
financial support of the Ministero della Universit\`a e della Ricerca
Scientifica e Tecnologica (MURST) under the program ``{Treatment of
large field astronomical images}'' and by the Agenzia Spaziale
Italiana.  PBS gratefully acknowledges the generosity of the
Universit\`a di Padova and the MURST for supporting his visit to the
Dipartimento di Astronomia.  
\end{acknowledgements}
\bibliographystyle{aabib99} 
\bibliography{mnemonic,biblio}
\end{document}